\documentclass{article}
\usepackage{arxiv}

\usepackage[utf8]{inputenc} 
\usepackage{url}            
\usepackage{booktabs}       
\usepackage{amsfonts}       
\usepackage{nicefrac}      
\usepackage{lipsum}
\usepackage{amsmath,amssymb}

\usepackage{changepage}

\usepackage{textcomp,marvosym}

\usepackage{cite}

\usepackage{nameref,hyperref}

\usepackage[right]{lineno}

\usepackage[nopatch=eqnum]{microtype}

\usepackage[table]{xcolor}

\usepackage{array}

\usepackage{enumitem}

\usepackage{tabularx}
\usepackage{caption}

\usepackage{subcaption}

\usepackage[T1]{fontenc}

\usepackage{multicol}

\usepackage{multirow}

\usepackage{amsthm}

\usepackage{algorithm2e}

\usepackage{lastpage,fancyhdr,graphicx}
\usepackage{epstopdf}

\usepackage{authblk}

\usepackage{titlesec}

\usepackage{float} 

\title{Herding Unmasked: Insights into Cryptocurrencies, Stocks and US ETFs}

\author[1,3,*]{An Pham Ngoc Nguyen}
\author[1,4]{Martin Crane}
\author[2]{Thomas Conlon}
\author[1,4]{Marija Bezbradica}

\affil[1]{School of Computing, Dublin City University, Whitehall, Dublin, Ireland}
\affil[2]{Smurfit Graduate School of Business, University College Dublin, Blackrock, Dublin, Ireland}
\affil[3]{SFI Centre for Research Training in Artificial Intelligence, Dublin, Ireland}
\affil[4]{ADAPT Center for Digital Content Technology, Dublin, Ireland}
\affil[*]{E-mail address: ngocan.nguyenpham6@mail.dcu.ie}

\begin{document}
\maketitle

\begin{abstract}
Herding behavior has become a familiar phenomenon to investors, with potential dangers of both undervaluing and overvaluing assets, while also threatening market stability. This study contributes to the literature on herding behavior by using a recent dataset, covering the most impactful events of recent years. To our knowledge, this is the first study examining herding behavior across three different types of investment vehicle and also the first study observing herding at a community (subset) level. Specifically, we first explore this phenomenon in each separate type of investment vehicle, namely stocks, US ETFs and cryptocurrencies, using the Cross-Sectional Absolute Deviation model. We find mostly similar herding patterns for stocks and US ETFs. Subsequently, the same experiment is implemented on a combination of all three investment vehicles. For a deeper investigation, we adopt graph-based techniques including the Minimum Spanning Tree and Louvain community detection to partition the combination into smaller subsets to detect herding behavior for each subset. We find that herding behavior exists at all times across all types of investment vehicle at a subset level, although perhaps not at the superset level, and that this herding behavior tends to stem from specific events that solely impact that subset of assets. Lastly, we explore herding by examining the financial contagion effects between these types of investment vehicle. Results show that  US ETFs not only have a tendency to propagate similar trading behaviors in stocks and especially cryptocurrencies but also show self-reinforcing herding behavior, acting as drivers of their own trends. 
\end{abstract}

\section{Introduction} \label{sec:1}

Market participants have long been known to play a crucial role in shaping the common price movements of financial assets~\cite{bik00}. High demand for an asset propels its prices upward, while a loss of interest or adverse news about the asset tends to lead to a price decrease. As a result, the understanding of investor reactions to various market conditions, e.g. financial crises, political events, bull and bear markets, is necessary to make wise investment decisions. Indeed, there have been a considerable number of studies focusing on herding behavior among investors in financial markets, from traditional ones like stocks and ETFs to alternative assets such as art real estate and cryptocurrency~\cite{amp23,tal21,ahn24,kin21,fre14,man22,alm23,kun21}. This remains a topic of interest as financial markets are constantly evolving and reacting to new events over time~\cite{amp23}. As defined by most studies, herding behavior describes collective actions from different investors, either purchasing or selling, targeting a set of different assets at a certain period which results in similar price movements between these assets~\cite{kin21}. This behavior typically arises from two scenarios~\cite{tal21}: First, intentional herding - where investors disregard their individual beliefs and follow the decisions made by the majority. Second, unintentional herding - where investors do not follow the market consensus and make a decision by themselves. However, in this case, they accidentally make the same decisions due to shared information sources such as financial reports, price momentum and analyses. It has been shown that the collective movements brought about by intentional and unintentional herding directly threaten market stability and are considered a key factor contributing to speculative bubbles~\cite{wang23}.

Different approaches have been proposed to detect herding signals. These can be classified into statistical and Neural Network methods. The former often involves constructing a regression model based on return values, then using optimization strategies such as Ordinary Least Squares (OLS) to obtain independent features' coefficients, which are assessed to identify the herding signals, using methods like  Cross-Sectional Standard Deviation (CSSD model)~\cite{chr95}, Cross-sectional Absolute Deviation (CSAD model)~\cite{cha20}, the Power Law Exponent-based model~\cite{ahn24} and the return-variance based model~\cite{kin21}. An alternative is to construct a metric based on a database of investors (e.g. the number of investors and demographics data)  and their trading activities (e.g. volume of purchasing and selling), then set a threshold for the metric to decide whether the herding appears. This approach needs more than the return values since a database of such trading activities is required, such as measures proposed by Lakonishok et al.~\cite{lak92}, Frey et al.~\cite{fre14} and Patterson et al.\cite{pat06}. Regarding Neural Networks, these methods are rather new in this area. Currently, one mainly deploys Deep Learning models such as Long Short Term Memory (LSTM)~\cite{fuj20} and BERT~\cite{ngu22} to gauge the market sentiment (e.g. positive, negative and neutral) which is then incorporated as an independent feature in regression models akin to statistical approaches~\cite{li23}. To this end, statistical approaches remain predominant in herding detection studies, with the CSAD model emerging as the most commonly utilized option~\cite{chi10,aha21,bou21,wak22,yan23}.

One finding common to most studies examining different asset classes is that investors tend to follow the market consensus during periods of economic uncertainty, which eventually results in herding behavior. Historical examples include events like Black Monday in 1987~\cite{lit16}, the Dot-Com/Tech crash in 2000~\cite{lit16}, the global financial crisis in 2007-2008~\cite{kab18}, and more recently the crash brought on by Covid-19 pandemic in 2020 ~\cite{man22}. Notably, it is worth mentioning that the role of algorithmic trading during market panics has been revealed, showing that this decision-making method still tends to operate during periods of high uncertainty~\cite{cha23}. However, this approach does not seem to reduce herding behavior. On the contrary, it appears to inadvertently cause herding due to identical decisions made by algorithmic trading systems~\cite{kle13}. In addition, studies in the cryptocurrency market indicate that its investors are more likely to behave irrationally and are easily influenced by price fluctuations as well as concrete items of news. In contrast, it has been suggested by a number of authors that professional investors tend to be dominant in traditional markets such as stocks, commodities and currencies. Besides, herding behavior in these markets is mainly explained by economic uncertainty as well as macroeconomic factors~\cite{alm23,kau23,ahn24,cho22}.   

Given the diversification of studies in herding behavior detection, most of them only focus on one particular type of investment vehicle. Since investors tend to diversify their portfolios by investing in different asset types~\cite{con24}, only considering one of them overlooks this investment manner, resulting in a shortage of insights. Our study addresses this limitation by using three popular types of investment vehicle, e.g. stocks, US ETFs and cryptocurrencies.

Furthermore, existing literature tends to employ a herding detection model on the entire dataset comprising all available assets. As a result, conclusions regarding the herding phenomenon are obtained based solely on the whole dataset, potentially yielding inadequate results. For instance, the herding might just be true for a portion of the dataset while leaving the others unaffected but the latter are still counted because they belong to the same dataset. We propose an approach to tackle this limitation. In particular, we first use Minimum Spanning Tree (MST) \cite{wkt21} and Louvain community detection~\cite{blo08} to group the assets based on their similarity in terms of price movements. Therefore, each group contains the assets that are most similar to each other. To recall, the fundamental idea of herding behavior is that assets display synchronized price movements due to the mutual trading actions of investors at the same time, especially during periods of high market uncertainty. Therefore, herding is more likely to occur within the assets that share similar price movements. By contrast, assets that are dissimilar to each other are less likely to demonstrate herding signals. Our way of clustering assets can help reduce false positives in the results since we assign assets with similar movement patterns to the herding detection models right at the beginning rather than a set of random assets with different price movements, which initially contradicts the fundamental nature of herding, as previously stated. Next, herding detection is applied to each group of similar assets. Following this practice, we can observe the herding behavior at a community level, offering clearer and more granular results compared to conducting herding detection on the entire dataset.

To address the shortcomings in the existing literature, our study aims to tackle the following research questions:
\begin{itemize}
    \item RQ1) Are herding patterns in each type of investment vehicle (namely stocks, US ETFs and cryptocurrencies) persistent, when considered separately? What causes the differences or similarities between the herding patterns?
    \item RQ2) By forming a combination of all assets across three investment vehicle types and splitting them into small communities based on their similarity, are there patterns in herding that manifest themselves in this combination at a community level?
\end{itemize}

To answer these questions, we utilize 30-min closing price time series of 222 assets, including 146 stocks, 49 US ETFs tracking their corresponding US indices and 27 cryptocurrencies. All of these data are sourced from the platform \href{https:\\Firstratedata.com}{FirstRate Data}. As mentioned earlier, the CSAD model is one of the most popular approaches for herding detection at the moment, thus, we also use this model in our present work for ease of comparison with others' work. Regarding community detection, we first construct an asset graph based on the Pearson correlation~\cite{coh09} between the assets, then apply Minimum Spanning Tree (MST)~\cite{wkt21} and Louvain~\cite{blo08} algorithms to the asset graph in order to identify the communities. The details of this process are described in Section \nameref{sec:3}. Lastly, and motivated by comments in the literature that herding and informational cascades on one asset result in similar effects on other assets~\cite{cip08}, we investigate whether such contagion effects manifest themselves among the three investment vehicles in our study via a well-known approach Vector Autoregression (VAR), marking a further novelty in the paper. 

Results obtained from our study can be summarised into the following core findings: Firstly, while herding patterns in stocks and US ETFs are mostly similar, there are differences between these and cryptocurrencies, meaning that these digital assets still remain separate from traditional assets in terms of herding behavior among their investors to some extent (RQ1). Secondly, by detecting herding behavior at a subset (community) level, we find that there are always subsets (communities) revealing herding behavior in each type of investment vehicle throughout the period examined, even though this phenomenon is not pronounced in the corresponding superset at some stages. Therefore, we conclude that while herding may not manifest when looking at all assets, it can still be found within specific subsets (RQ2). Thirdly, we notice 4 stock sectors (i.e. Information Technology, Financials, Healthcare and Energy), wherein assets within each sector and their corresponding ETFs tend to form communities themselves. Furthermore, the herding behavior observed in these communities tends to stem from specific events that solely impact that sector (RQ2). Fourthly, US ETFs reveal strong contagion effects to both stocks and cryptocurrencies, especially during times of market uncertainties, indicating their potential to be drivers of herding in financial markets. On the contrary, although stocks and cryptocurrencies also exhibit evidence of shock transmission to US ETFs, these effects are quite weak. Instead, US ETFs display an internal contagion phenomenon by transmitting shocks to themselves, potentially leading to a self-reinforcing herding behavior.

The remainder of the article is organized as follows: Section \nameref{sec:2} goes through related works. Section \nameref{sec:3} provides a description of the datasets followed by methodologies. Section \nameref{sec:4} describes the experimental results followed by implications and hypothesis. Lastly, the conclusion of this study is given in Section \nameref{sec:5}.

\section{Related Works} \label{sec:2}
\subsection{Periods of Uncertainty in Recent Years}
Throughout the 4-year span from 2019 to 2023, several significant events have hit global financial markets, including the trade war between China and the US~\cite{mol23}, the outbreak of the Covid-19 pandemic~\cite{cio20}, and, more recently, the political war between Russia and Ukraine~\cite{ass23}. While these events arose for different reasons, they collectively caused negative influences on financial markets, affecting both traditional and cryptocurrency assets to some degree.

The US-China trade war took place in January 2018, following the imposition of initial tariffs on Chinese imports by the new US administration at that time~\cite{mol23}. The primary goals behind this decision were to reduce the US trade deficits, stop the requirements of American companies to transfer technology and protect US technology companies from intellectual property theft~\cite {su24}.  As a countermeasure, China also levied different tariffs on the US products~\cite{mol23}. Although the trade war started early in 2018, it escalated during 2019~\cite{apc21}. Consequently, financial markets in both the US and China were significantly affected, especially in the commodity, energy and technology sectors~\cite{che23,bai23,apc21}. Major stocks in both countries as well as the ETFs that track them fell sharply~\cite{kap20,ric19}. Trade war headlines attracted huge attention throughout the year. Announcements relating to the tariff imposition caused the stock and index values to fluctuate dramatically~\cite{lew19}. Indeed, it has been shown that there was a wave of selling pressure in 2019 because the stock prices fluctuated due to the panic of investors caused by the trade war~\cite{par19}.  In China, the economy also exhibited a decline in GDP growth rate in 2019~\cite{su24}, accompanied by decreased returns across various sectors in financial markets~\cite{chen22}. The uncertainty in financial markets due to this trade war was also seen in other countries across the world~\cite{ben23,bai23}. The event ended in January 2020 after both countries signed off on the so-called \textit{Phase One agreement} trade deal~\cite{bow21}.

Right after the trade war, the world encountered another challenge in the form of the Covid-19 pandemic~\cite{cio20}. This event disrupted economies across the world. In particular, due to the fast spread speed and serious consequence of this health crisis, different regulations were applied to stop the virus, such as closing national borders, constraining mobility as well as stopping business activities worldwide~\cite{mar21,bar20}. Subsequently, these measures profoundly influenced the economies of numerous countries, eventually turning into a global economic downturn in March 2020 as a response to various government’s efforts to control the disease spreading~\cite{pad21}.  This is considered one of the biggest financial crises ever, even worse than previous ones, e.g Black Monday 1987, 2007-2008 global financial crisis~\cite{li22}. Various studies highlighted that this pandemic has disturbed and devalued financial markets, from traditional assets such as stocks and bonds~\cite{lah20,you21,ach21} to safe-haven assets such as gold~\cite{ali20,che21} to the cryptocurrencies~\cite{lah20,dro20,ngu23}. The pandemic's strongest effects were observed during its first wave (from March until June 2020), with most assets experiencing a big drop in prices~\cite{pha22}. Fortunately, although the world continued facing different Covid-19 waves, its impact on financial markets lessened significantly, with their prices gradually recovering to pre-pandemic levels~\cite{cao21}.

As the impacts of Covid-19 were being reduced through intervention measures such as vaccines and social distancing, another major event disrupted world economies. On 24 February 2022, Russia officially declared war on Ukraine. Numerous studies have dug into the impact of this event on the financial markets~\cite{der23,das23,conlon24,kum23,gai22,bou22}. Russia, a significant player in oil and gas production, has long been a crucial fuel supplier to many countries, particularly in the EU. Furthermore, Ukraine has served as a critical transit route for Russian petroleum exports to Europe. Consequently, this conflict had a severe effect on the energy sector~\cite{tee23}. In particular, when tensions rose and military actions escalated, oil prices tended to increase due to fears of potential supply disruptions. Conversely, when diplomatic efforts appeared to make progress or when there was a decrease in hostilities, oil prices tended to ease~\cite{mag24,hao23}. Moreover, increased sanctions on Russia and efforts by countries, notably in Europe, to reduce dependence on Russian energy further triggered oil and gas shortages, the energy prices in Europe thereby surged during the first half of 2022 until alternative suppliers were found later on~\cite{bab23,rok23}. Additionally, the EU's embargo on seaborne crude oil and the G7's price cap mechanism, which both took effect on December 5, 2022 also pressured the global oil prices already high for seasonal reasons~\cite{bab23}. Regarding other sectors, most studies have focused on the EU financial markets, since they were directly linked to Russia and Ukraine in commercial and political activities. Generally, it has been shown that the war had heterogeneous impacts on the EU financial markets. In particular, while most countries revealed a drop in their stock market prices after the war took place~\cite{kum23,ahm23,gai22}, those closest to Russia were impacted greater~\cite{boun22}. Interestingly, stocks and indices in several countries experienced an upward trend during the war, such as Belgium, Denmark, Luxembourg, Netherlands and Spain~\cite{kum23}. On the other hand, this war seemed not to have distorted the financial markets in Asian countries and other developed ones such as the US~\cite{bou22,ass23,imr22}. Nevertheless, studies~\cite{wu23,imr22,gai22} emphasized that, aside from the energy sector, the effect of this event on financial markets globally is negligible compared to previous events such as the Covid-19 pandemic, the trade war and past financial crises.

\subsection{Herding Behavior in Traditional Markets}

The discussion of herding behavior among investors dates back over a century, paralleling the long historical development of traditional markets~\cite{sch36}. One of the earliest topics discovered in this area deals with herding behavior under different market conditions. In~\cite{aha21}, Aharon applied 2 popular herding detection methods, CSSD and CSAD, to a dataset comprising historical daily data spanning from 1990 to 2019. This dataset consists of 10 size-ranked portfolios constructed from stocks traded on the NYSE, AMEX and NASDAQ, covering different major events like the Asian crisis, the Dot-Com bubble, and the 2007-2008 global financial crisis. He found evidence of herding within these stock assets throughout the examined period. Moreover, by integrating the volatility index (VIX index) into the herding models, he noticed that herding signals intensified when market volatility - as measured by the VIX index - increased. The same conclusion was also reached by Ahn et al~\cite{ahn24} who used different datasets and herding detection methods. Specifically, based on the daily US stock return time series of 137 firms listed in the Standard \& Poor 500 (S\&P500) from January 1992 to December 2021, they utilized the Power law exponent-based model to calculate the time-varying herding magnitude and then compared the results with an economic uncertainty index adopted from \cite{blo14}. With access to longer and more recent data, the authors confirmed that, apart from the major events covered in the above study,  the herding was also more pronounced during the Covid-19 pandemic.  

However, one downside of these studies is that they detected the herding over an entire period, leading to a lack of details. Thankfully, several studies filled this gap by using shorter periods at different points in time, creating a more complete picture of this phenomenon. For example, Chiang and Zheng in~\cite{chi10} observed the herding of 63 stocks originally from Thailand between  January 1997 and December 1998 - a period that covers the Asian financial crisis. Using the CSAD model and daily return time series, they suggested the presence of herding in these stocks during the crisis. On the other hand, in \cite{guv16}, the authors collected monthly prices from the 20 most traded stocks in the Egyptian market over 12 years from July 2002 until May 2014. As distinct from most studies, the authors used a state-space model to detect the herding while dividing the data sample into pre-crisis, during the 2007-2008 crisis and the post-crisis period. Their findings revealed an absence of herding behavior before the crisis, transitioning to herding during the financial crisis and persisting afterward.   

Notably, with the huge effect of the Covid-19 pandemic, many studies have explored herding behavior during this unprecedented time~\cite{bou21,has23,yan23,asl22,tal21,zhu24}. In~\cite{bou21}, Bouri et al. selected 49 daily stock market indices representing 49 different countries sourced from Morgan Stanley Capital International (MSCI) indices. Using the well-known CSAD model and a rolling window with a size of 102 days, they examined the time-varying global herding from  January 2019 until August 2020. Their idea was to assess whether financial markets around the world exhibited the herding phenomenon during the onset of the pandemic. Indeed, it was shown that herding was apparent from March to May 2020, coinciding with the worst period of the global economy caused by the pandemic. Other studies focusing on different types of investment vehicle, timeframes and methodologies also obtained the same results as investors tended to mimic each other's investment decisions during the first wave of the pandemic (first and second quarter of 2020)~\cite{hua24,amp23}. 

Another aspect explored in the study of herding behavior is the comparison between different countries. A work by \cite{mou24} conducted a literature review about herding behavior in different countries from 1995 to 2021, its authors found differences in external factors driving herding within different regions. For instance, global risk factors were identified as primary culprits influencing the stock market and giving rise to herding behavior among investors in the Turkish market~\cite{bal15}. On the other hand, the Chinese market tended to be affected by factors such as analyst recommendations, short-term investors' horizons as well as systemic risk~\cite{cho17}. Moreover, the authors also noticed that the majority of markets in developing countries were exposed to herding among retail investors (i.e. non-professional traders relying on personal judgment and subjectivity to make investment decisions). The purchasing power of these investors tends to be relatively small as limited by their personal earning ability. On the other hand, retail investors in developed countries tended to display less herding behavior due to the greater availability of information and knowledge of stock prices. Another study conducted by Chen and Zheng~\cite{che22} employed the Simulated Method of Moments estimator to investigate the herding behavior in the Chinese and US stock markets. The experiment was based on the Shanghai Composite Index representing the China market and the S\&P500 Index representing the US market over the period between 01/01/1992 and 10/30/2017, including 6739 daily observations. The authors revealed both similar and different patterns in herding behavior between the two countries. Furthermore, they stated that the Chinese stock market was mainly driven by behavioral sentiment dynamics due to the switching behaviors of investors while the US market was mainly driven by fundamental factors. The authors explained that this contradiction was attributed to the lack of a sufficiently transparent environment as well as comparatively few sophisticated investors in China, as compared to well-developed markets like the US. Additionally, over-speculative trading and the weak determining force of fundamental factors in China further contributed to this phenomenon. More recently, Yang and Chuang in~\cite{yan23} investigated herding behavior across three different countries, including China, Taiwan and the US. They collected stock returns for non-financial firms listed on the Taiwan Stock Exchange (TWSE), the Shenzhen Stock Exchange (SZSE), the Shanghai Stock Exchange (SHSE), and the New York Stock Exchange (NYSE) from  January 2001 until June 2021. To observe the herding behavior, they combined the CSAD model with the Kalman filter and GARCH models. The authors discovered a relatively similar result to the above study. Specifically, Taiwan and China exhibited identical herding patterns with significant herding observed in two sub-periods: 2001-2004, coinciding with two local events, namely the Taiwan Strait crisis and the SARS pandemic and 2007-2009, corresponding to the global financial crisis. By contrast, although herding was also pronounced during 2007-2009 in the US, no herding was apparent before this sub-period. Instead, this developed market revealed another instance of herding during the Covid-19 pandemic in 2020.

In general, most studies in traditional markets found herding displays during events that disrupt financial and economic stability. As explained in~\cite{tal21,zhu24,gho23}, during such events, investors tend to abandon their personal information and are influenced by financial news as well as the actions of other investors. Moreover, investors in traditional markets tend to be easily influenced by word of mouth, given the increase in uncertainty and negative news surging, especially retail investors. Additionally, during uncertain times, both retail and institutional investors behave similarly and noise trading is shown to increase during the high volatility. On the other hand, institutional investors are dominant when the markets experience an upward trend, resulting in asymmetric herding during the positive market condition~\cite{ahn24}.

\subsection{Herding Behavior in the Cryptocurrency Market}

Cryptocurrencies are a relatively new asset class compared to traditional assets. Although the concept of digital coins was introduced in 2009 with the emergence of Bitcoin as the pioneering cryptocurrency~\cite{lut16}, trading activities in this market only really commenced in 2013~\cite{mai20}. As a result, studies on herding behavior in this domain mainly focus on the period from 2013 onward. In~\cite{poy18}, Poyser identified herding in the top 100 cryptocurrencies by market capitalization from late April 2013 until mid-October 2019. He first used a Markov-switching model to segment the sample data into different regimes, then relied on a CSAD model to detect the time-varying herding in each regime. He found that herding behavior changed over time, with evidence of both herding and its absence throughout the examined period. Notably, herders were more active from mid-2017 until the end of the period since the signal of herding intensified during this interval. Additionally, in the second experiment, the author distinguished between 2 market conditions, namely positive and negative returns, to observe the herding behavior in each situation. The results revealed that cryptocurrency investors were more likely to follow the consensus when the market exhibited positive returns, while this tendency was diminished during periods of negative returns. Similarly, a study conducted by Raimundo et al.~\cite{rai22} collected more recent data, including 80 cryptocurrencies with the highest market capitalization from \href{https://coinmarketcap.com/}{coinmarketcap.com} between July 2015 and March 2020. In this work, they applied a state-space model adapting the standardized-beta method to detect the herding and used a regression model to test the relationship between the herding behavior and the market volatility. Like the previous study, the authors indicated a mix of herding and no herding throughout the considered period. However, they revealed a non-herding signal during most of 2019. Besides, they also found that the herding in cryptocurrencies became more pronounced when the market volatility increased. This finding tallies with various studies, using different datasets and methodologies~\cite{man22,jal20}.

Similar to traditional markets, the Covid-19 pandemic has also attracted significant attention in the cryptocurrency market. In~\cite{man22}, the authors gathered intraday trading data for 9 large cryptocurrencies such as Bitcoin, Ethereum and Litecoin, covering a 2-year period from January 2019 to January 2021. They applied  Patterson and Sharma's herding intensity measure to each cryptocurrency in two distinct sub-periods: Pre-Covid-19 (January 2019 - December 2019) and Covid-19 pandemic (January 2020 - January 2021). Unlike other aforementioned approaches, this method can detect the herding activities in each asset. Results suggested a dramatically greater herding intensity during the latter sub-period across all examined cryptocurrencies. Nevertheless, while all cryptocurrencies showed the herding phenomenon during this time, no evidence of herding was found in Bitcoin, Ethereum, and Ripple during the Pre-Covid-19 sub-period. These findings align with those of Youssef and Waked who executed a hybrid model combining CSAD and Markov-switching methods on 43 cryptocurrency price time series with daily frequency~\cite{wak22}. In particular, their dataset spans from 28/04/2013 to 11/11/2020, divided into two sub-samples based on the Covid-19 pandemic, including the Pre-Covid-19 from the beginning until the end of 2019 and the Covid-19 outbreak thereafter. As expected, only the second sample (i.e. the Covid-19 outbreak) revealed the herding evidence. Furthermore, for the second part of the work, the authors used the coronavirus media coverage index (MCI) which calculates the percentage of all news sources covering the novel coronavirus topic to test the relationship between news and the herding magnitude. Interestingly, they found that the news about the pandemic significantly drove the herding behavior among investors in the cryptocurrency market, indicating the fact that cryptocurrency traders are mainly influenced by news and people's sentiments. The result is in line with most of the existing literature~\cite{vid18,har20,bur21}.

More recently, a few studies have utilized a newer dataset that covers up to the year 2023~\cite{dia23,wan23}. However, instead of segmenting the sample into smaller intervals like the previous studies did, they mainly retained the whole period which caused ambiguity for herding detection models.  Consequently, their findings tended to suggest an absence of herding throughout the examined period, even during highly uncertain times such as the outbreak of the Covid-19 pandemic. Additionally, most studies investigating the financial markets (both traditional and cryptocurrency) focus on periods before the year 2022. Given these current limitations, our work contributes to the existing literature by using more up-to-date data, which spans from April 2019 until May 2023. Moreover, we segment our data into finer sub-periods corresponding to major events that occurred in the financial markets throughout the examined period. Further details of this time division will be described in Section \nameref{sec:3}.


\subsection{Contagion Effects in Financial Markets}
In the study of financial contagion, understanding how shocks propagate across assets and markets is critical to assessing systemic risk and interdependencies. Financial contagion refers to the transmission of economic shocks between distinct markets or assets, which often leads to heightened volatility and amplified co-movement across financial instruments.Herding and Contagion are frequently considered together in the financial literature, especially in the context of cryptocurrencies \cite{Bukhari_Ahmad_Hanif_2021}, \cite{Kumar2020}.Given evidence of herding within a set of assets, the primary transmitter of shocks can be identified through observed contagion effects. For instance, \cite{dag19} examined herding behavior and contagion effects among 50 cryptocurrencies over a four-year period from 2015 to 2018. The authors found a weak herding signal within this group of cryptocurrencies, with Bitcoin being the main herding driver, as evidenced by its significant contagion effect on other coins.

Regarding contagion between asset classes, a number of authors have investigated such risks between cryptocurrencies and stock markets and a selection of these is described below.
In \cite{MATKOVSKYY201993}, the topic of contagion from stock markets to Bitcoin (BTC) futures after BTC futures were launched is considered, indicating increased evidence of this after the launch. The authors further note that risk averse investors tend to avoid Bitcoin markets, preferring mature markets such as NASDAQ and NIKKEI, after the launch of BTC futures and in bear market periods.
The authors in \cite{CAFERRA2021101954} compared cryptocurrency and stock market dynamics during Covid-19 and discussed co-movement and hidden regimes of cryptocurrencies and stocks in Covid-19 using the wavelet coherence and Markov switching approaches. The authors found that both markets correlate over time at low but not at high frequencies.  Another study, \cite{WANG2022102345} examined asymmetric contagion effects between stock index data and the largest two cryptocurrencies (ETH and BTC) using a time-varying symmetrized Joe-Clayton copula GARCH model with a nonlinear Granger causality test for data between 2017 and 2020. They found evidence of time-varying tail dependence, with lower tail dependence found to be greater relative to upper and asymmetric contagion between the stock and cryptocurrency markets.  
A recent paper \cite{Niyitegeka2023}
looks at contagion between cryptocurrencies and stocks in emerging and mature markets during the Covid-19 period with an approach using DCC Garch and Wavelet Cross-Correlation.  The data featured BTC as a proxy for cryptocurrencies, MSCI on behalf of emerging markets and S\&P500 closing prices from October 2014 to March 2022. Their findings show that during periods of financial upheaval (compared to more tranquil ones) an increase in conditional correlation was manifest. Evidence from wavelet cross-correlation analysis showed a positive cross-correlation between the BTC and equity markets in periods of financial unrest, with evidence for cross-correlation found at fine and coarse scales.
Another recent paper dealing with a similar topic \cite{dong2023tracing} found the cryptocurrency-stock correlation to be largely absent pre-March '20 but increased considerably after. The data examined in the paper was unusual in that intraday as well as daily data were examined for BTC from July 2015 to June 2022.  For stock indices during this period, SPY, DIA and QQQ were used as proxies for S\&P500, Dow Jones, and Nasdaq, respectively. The authors speculated that the behavioral shift found was fueled by the Federal Reserve’s policy pandemic policy response to the Covid-19 pandemic.  Also, little evidence of cryptocurrency shocks to stocks but large volatility spillovers in reverse were elicited. Finally, the authors linked increases in cryptocurrency-stock co-movement after the COVID-19 pandemic to the increased numbers of institutional investors in the cryptocurrency markets, where trades are sensitive to monetary policy changes.

In terms of contagion between stocks and ETFs, Da and Shive \cite{da2018exchange} saw that both information and non-fundamental shocks can be transmitted by ETFs. 
 This takes place through an increased demand for ETFs resulting in price pressure, with onward transmission to the underlying basket of shares.  The authors attribute this to arbitrageurs simultaneously taking opposite positions in the ETF and the underlying shares, with this effect being particularly noticeable for small and illiquid stocks.
\cite{wang4904298idiosyncratic}
 took a network approach to examining risk spillover effects between ETFs and stocks. He looked at ETF-stock spillovers in different industrial sectors and also found - for certain sectors - a tendency for ETFs' shocks to be transmitted to their constituent stocks. In addition, he commented on periods of high market volatility, when the contagion source tends to diversify.  This effect is also noted in \cite{NING2024103194} who found, in addition, that ETFs also propagate tail event shocks from one stock to their other constituent stocks as well as propagate demand shocks. Gebka and Wohar \cite{akb13} noted different price patterns in different sectors which they attributed to traders' irrationality with this being particularly the case for basic materials, oil and gas and consumer services. The authors suggested that this phenomenon can be driven by investors moving in and out of markets in sync, an extreme preference for safe assets or overconfidence.

\section{Methodology and Data Collection} 
\label{sec:3}
\subsection{Data Sources}
The historical time series utilized in this study are collected from the \href{https://firstratedata.com/}{FirstRate Data}, a leading provider offering high-resolution intraday prices of financial assets on a variety of asset classes. Our dataset comprises 222 assets categorized into 3 types of investment vehicle, namely cryptocurrency (27 assets), stock (146 assets) and the US ETF (49 assets). The selection of these assets is based on four criteria: 
\begin{enumerate}[label=\roman*.]
    \item Data availability: the assets available on the FirstRate platform.
    \item Market capitalization: since we are limited to little more than 50 assets for cryptocurrencies and US ETFs, we try to use all of them. On the other hand, there are thousands of stocks available to use so we narrow these down to the top 200 stocks whose companies are in the top 200 in terms of market capitalization, according to the web page \href{https://companiesmarketcap.com/}{Companies Market Cap}.
    \item High trading frequency: To avoid missing values, we only use assets with high trading frequency. As a result, we have to remove the assets that have more than 1, 10 and 12 percent of missing values among stocks, cryptocurrencies and US ETFs, respectively. Indeed, after filtering the assets, the majority of the remaining have less than 1 percent of missing values.
    \item In the case of US ETFs: the ability to act as a proxy for a corresponding US stock index.
\end{enumerate}

Each time series covers the period from 01/04/2019 until 03/05/2023 and has a resolution of 30 minutes. Since the official trading time of traditional markets is between  09:30 a.m. and 4:00 p.m., we select this timeframe as the boundary for our time series, meaning that the cryptocurrency data is only available within this trading time, even though these coins are traded nonstop. Other price values outside of this timeframe are not considered in our experiments. This practice facilitates the comparison between assets from different classes and also preserves their equality. A list of assets in each class is given in \nameref{S1_Tab}.

\subsection{Financial Sector Classification}
Unlike cryptocurrencies and US ETFs, the stock market is classified into 11 sectors based on their respective business activities: Communication Services, Utilities, Real Estate, Materials, Information Technology, Industrials, Healthcare, Financials, Energy, Consumer Staples, and Consumer Discretionary. In addition, while there is no official classification for ETFs and most contain assets from different stock sectors, we notice the existence of ETFs designed to track the performance of a specific stock sector. For instance, ticker IYW represents the iShares U.S. Technology ETF which tracks technology stocks. Thus, although we do not classify ETFs, we keep in mind this characteristic. As a result, we assign each asset to a specific sector, either cryptocurrency, US ETF, or one of the aforementioned stock sectors. We use this information in our experiments for a deeper investigation and analysis of the grouping of such assets.

We note that the number of assets in each stock sector is different and depends on its asset's market capitalization. For instance, Technology, Financials and Healthcare stocks tend to have very high market capitalization while Communication Services and Real Estate tend to have lower market capitalization~\cite{dal24}. Consequently, among the 11 stock sectors, Utilities, Materials, Real Estate and Communication Services exhibited a limited number of entities. Therefore, this study concentrates only on the remaining 7 sectors. Further details for this classification are provided in \nameref{S2_Tab}.

\subsection{Graph Construction and Minimum Spanning Tree}

A graph is a set of objects (referred to as \textit{nodes}) that show connections with each other. That is, if there is a connection between two objects, it is marked by a link (referred to as an \textit{edge}) connecting them together. The definition of this connection depends on the purpose of the graph. For instance, ~\cite{cuo24} constructed an electric bicycle-based commute graph whose nodes represent locations and the connection between two nodes is measured by the number of trips traveling between the two corresponding locations that use this kind of bicycle. On the other hand, the Belgian mobile phone graph was introduced in~\cite{lam08} where the authors examined the connections among phone users via their phone calls and text messages' frequency. Regarding financial markets, a graph between financial assets tends to focus on the correlation between them, i.e. the similarity in price movements between the assets~\cite{wkt24,dro20,pha22,ngu23}. In this study, our graph is constructed as follows.

Given a price time series $x_i = (x^1_i, x^2_i, \dots, x^T_i)$ where $i$ represents an asset, $T$ represents the length of the time series, $x^t_i$ represents the price value of asset $i$ at time $t$. The log-return at time $t$ of asset $i$ is defined as  $r^t_i = log(r^{t}_i/r^{t-1}_i)$. Although the price values are undoubtedly informative, we use the return value for this study as it is represented as a percentage, making them scale-free and especially, stationary, which is an important requirement for statistical tools, such as herding detection models~\cite{tsa05}.

To obtain a graph of assets, we first construct a correlation matrix $\mathbf{C}$ based on the similarity between the assets. In particular, each element $c_{ij}$ in $\mathbf{C}$ represents the similarity between 2 assets $x_i$ and $x_j$ such that  
\begin{equation}
   c_{ij}  = \frac{\sum\limits_{t = 1}^T(x_i^t - \Bar{x_i})(x_j^t - \Bar{x_j})}{\sqrt{\sum\limits_{t=1}^T(x_i^t - \Bar{x_i})^2 \sum\limits_{t=1}^T (x_j^t - \Bar{x_j})^2}} \label{eq:1}  
\end{equation}

Where $\Bar{x_i}, \Bar{x_j}$ are the average values of time series $x_i, x_j$, respectively. This similarity measure is known as the Pearson correlation measure~\cite{coh09} and ranges from -1 to 1, where -1 indicates a total negative linear relationship, i.e. two time series are completely different, 0 indicates no linear correlation, and 1 indicates a total linear relationship, i.e. two time series are completely similar.

We then determine a distance matrix $\mathbf{D}$ where each element $d_{ij} = \sqrt{2(1 - c_{ij})}$. This distance matrix transformation is necessary as it satisfies the characteristics of a metric~\cite{sea06} that are necessary for further steps. Essentially, $d_{ij}$ represents the distance between two time series $x_i$ and $x_j$, ranging from 0 to 2 in which 0 indicates a complete similarity and 2 indicates a complete difference between them. 

From the distance matrix, we construct an asset graph where each node is an asset and an edge between two nodes depicts the relationship between them, which is the similarity calculated in the previous steps. One downside of the graph is that it is dense, making it memory-intensive and time-consuming for processing because all nodes are linked together while there are edges with negligible information. As a result, it costs a considerable amount of time to process the graph and redundant information might confuse experimental results. Thus, we extract a subset of the graph called \textit{Minimum Spanning Tree} (MST)\cite{wkt21} to reduce the graph size and remove the redundant information. This subset keeps edges with the highest correlation degree, guarantees the characteristics of a tree and ensures other edges are removed. Hence, only the important part of a graph is kept. As a result, this technique reduces our graph size of  222 nodes and 24531 edges to 222 nodes and 221 edges.

\subsection{Louvain Community Detection}

Community detection aims to partition nodes of a graph into separate groups (called \textit{communities}) by maximizing the similarity between nodes within a community while maximizing the distinction between different communities. In other words, this detection results in communities where nodes belonging to a community are most similar to each other while nodes from different communities exhibit different characteristics. In the context of financial markets, where nodes represent financial assets, a community in this case comprises assets that have most similar price movements over a certain period of time. On the other hand, assets from different communities exhibit less similar price movement patterns. For this technique, Louvain community detection algorithm is well-known for its applications in large networks~\cite{blo08}. Additionally, this approach is rather straightforward and is also adopted widely in various research topics~\cite{dib24,mai22,mai23,yuh24}. Since our graph has more than 200 nodes, this method is unquestionably well-suited for this case.

The Louvain algorithm is based on modularity optimization and its fundamental principle is searching for a community partition such that it maximizes the density of edges inside communities in comparison with edges between different communities. This optimization is executed through maximizing a measure called \textit{modularity}:

\begin{equation}
    Q = \frac{1}{2m} \sum\limits_{i,j}\left[A_{ij} - \frac{k_i k_j}{2m}\right]\gamma\left(g_i,g_j\right)\label{eq:2}  
\end{equation}

Where $A$ represents a graph, $A_{ij} = 1$ if there is an edge between node $i$ and node $j$ and 0 otherwise, $m = \frac{1}{2}\sum\limits_{i,j}A_{ij}$ is the total number of edges in the graph, $k_i = \sum\limits_{j}A_{ij}$ is the node degree of node $i$, $g_i$ represents the community where node $i$ belongs to, $\gamma\left(g_i,g_j\right) = 1$ if $g_i = g_j$ (nodes $i$ and $j$ belong to the same community) and 0 otherwise. To this end, the best community partition is obtained when $Q$ is maximum.

From this idea, Louvain community detection algorithm comprises two phases, as shown in Pseudocode \ref{pse:1}. Furthermore, the whole process of constructing communities from a set of financial time series is illustrated and interpreted in \nameref{SI_Appendix1}.

\begin{algorithm}[!h]
\caption{Louvain Community Detection}
\label{pse:1}
\KwIn{Graph $A = (V, E)$ with $V$ as the set of nodes and $E$ as the set of edges.}
\KwOut{Set of communities}
\textbf{Initialize:} Each node in the graph is assigned to a separate community.

\Repeat{No further improvement in modularity}{
  \textbf{Phase 1}\;
  \While{modularity $Q$ can be improved}{
    \For{each node $i \in V$}{
      Calculate the modularity gain $\Delta Q$ in turn for moving $i$ from its current community to each neighboring community\;
      \[
      \Delta Q = \left[ \frac{\sum_{in} + 2k_{i,in}}{2m} - \left(\frac{\sum_{out} + k_i}{2m}\right)^2\right] - \left[\frac{\sum_{in}}{2m} -\left(\frac{\sum_{out}}{2m}\right)^2 -\left(\frac{k_i}{2m}\right)^2\right]
      \]
      Where $\sum_{in}$ is the number of edges between nodes inside the community $g_j$, $\sum_{out}$ is the number of edges linked to nodes inside $g_j$ (including edges from nodes within other communities), $k_{i, in}$ is the number of edges from node $i$ to nodes in $g_j$\;
      \eIf{$\Delta Q$ in all cases is not positive}{
        $i$ stays in its current community\;
      }{
        Move $i$ to the community where $\Delta Q$ is positive and maximum\;
      }
    }
  }
  \eIf{modularity $Q$ after Phase 1 increases}{
    \textbf{Phase 2}\;
    Construct a new graph $A'$ where each community found in Phase 1 is considered as a node of $A'$\;
    The weight of an edge between two new nodes is the number of edges between nodes in the two corresponding communities\;
    $A \leftarrow A'$\;
  }{
    Stop the algorithm\;
  }
}
\end{algorithm}

\subsection{Herding Detection Strategies}
Cross-Sectional Absolute Deviation (CSAD) model is one of the most well-known herding behavior detection approaches that has been used widely in many studies\cite{jal20,you22,dia23,bou21,gho23}. The fundamental idea of this technique is to examine the relationship between the average return (market return) of a set of assets and the dispersion across each asset return (CSAD). A herding signal is pronounced when the market return changes significantly (i.e. the market experiences large fluctuations, either going up or down) while the dispersion between asset returns decreases (i.e. the return movements of different assets become more similar). In this study, we use two forms of the CSAD model for our experiments.
\begin{equation}
    CSAD_t = \beta_0 + \beta_1|r_{m,t}| + \beta_2 r_{m,t}^2 + \epsilon_t \label{eq:4}
\end{equation}

\begin{equation}
    CSAD_t = \gamma_0 + \gamma_1 D^{up} |r_{m,t}| + \gamma_2 D^{up} r_{m,t}^2 + \gamma_3 D^{down} |r_{m,t}| + \gamma_4 D^{down} r_{m,t}^2 + \zeta_t
    \label{eq:5}
\end{equation}

Where $CSAD_t = \sum\limits_{k = 1}^T |r_{i,t} - r_{m,t}| / T$; $r_{i,t}$ is the return of asset $i$ at time $t$; $r_{m,t} = \sum\limits_{k = 1}^{N} r_{k,t} / N $ is the average of asset returns (also known as the market return) at time $t$; $\beta_0, \beta_1, \beta_2, \gamma_0, \gamma_1, \gamma_2, \gamma_3, \gamma_4$ are regression coefficients, $D^{up} = 1$ if $r_{m,t}$ is positive, otherwise $0$; $D^{down} = 1$ if $r_{m,t}$ is negative, otherwise $0$; $\epsilon_t, \zeta_t$ are error terms at time $t$.

Regarding Eq \ref{eq:4}, a herding exists when the parameter $\beta_2$ is significantly negative (a negative coefficient with a confidence level of at least 90\%). Intuitively, we expect this coefficient to be negative because it follows the idea of the herding behavior, i.e. a dramatic increase or decrease in the market return ($r_{m,t}^2$) coincides with a decline (negative $\beta_2$) in the dispersion between corresponding assets' returns ($CSAD_t$). Eq \ref{eq:5} refers to a more detailed version of Eq \ref{eq:4} in which we separate the market condition into \textit{up} (i.e. positive market return) and \textit{down} (i.e. non-positive market return) so that the herding can be detected during the \textit{up} and \textit{down} market condition separately, instead of considering all market conditions (i.e. both \textit{up} and \textit{down} equally) as declared in the first form. The use of both forms for herding detection comes from an expectation that there might be periods when herding does not exist when considering both up and down market conditions at the same time but it might be found in the \textit{up} or \textit{down} market condition only. To this end, the same logic of Eq \ref{eq:4} is applied to Eq \ref{eq:5} wherein a significantly negative $\gamma_2$ refers to a herding signal during the \textit{up} market condition while a significantly negative $\gamma_4$ refers to a herding signal during the \textit{down} market condition. 

\subsection{Contagion Detection Strategy}
To capture the contagion effects between multiple time series, the Vector Autoregression (VAR) model has emerged as a foundational econometric tool. The VAR model, first introduced by Sims \cite{sims80}, is a statistical model that captures the linear interdependencies among multiple time series by allowing each variable to be regressed on lagged values of itself and other variables in the system~\cite{lon10}. Therefore, this model can identify bidirectional shock transmission between any two time series, capturing shocks moving from one series to the other and vice versa. Specifically, given a set of $k$ time series $\{y_1,y_2,..., y_k\}$, the VAR model of order $p$ to assess the contagion effects between these $k$ time series can be expressed as:

\begin{equation}
    \begin{bmatrix}
y_{1,t} \\
y_{2,t} \\
\vdots \\
y_{k,t}
\end{bmatrix} =  \begin{bmatrix}
c_1 \\
c_2 \\
\vdots \\
c_k
\end{bmatrix} + \begin{bmatrix}
a_{11}^{t-1} & \cdots & a_{1k}^{t-1} \\
a_{21}^{t-1} & \cdots & a_{2k}^{t-1} \\
 & \vdots & \\
a_{k1}^{t-1} & \cdots & a_{kk}^{t-1}
\end{bmatrix}
\begin{bmatrix}
y_{1, t-1} \\
y_{2, t-1} \\
\vdots \\
y_{k, t-1}
\end{bmatrix} + \dots + \begin{bmatrix}
a_{11}^{t-p} & \cdots & a_{1k}^{t-p} \\
a_{21}^{t-p} & \cdots & a_{2k}^{t-p} \\
 & \vdots & \\
a_{k1}^{t-p} & \cdots & a_{kk}^{t-p}
\end{bmatrix}
\begin{bmatrix}
y_{1, t-p} \\
y_{2, t-p} \\
\vdots \\
y_{k, t-p}
\end{bmatrix}
+
\begin{bmatrix}
\varepsilon_1 \\
\varepsilon_2 \\
\vdots \\
\varepsilon_k
\end{bmatrix}\label{eq:6}
\end{equation}

\noindent where $y_{i,t}$ is the value of time series $y_i$ at time $t$; $p$ is the number of lagged values of each predictor time series used to predict the current value of a response time series. In other words, each time series in the system depends not only on its own past values up to $p$ time steps but also on the past values of all other time series in the system up to $p$ time steps; $c_i$ is a regression coefficient for the corresponding equation of time series $y_i$; $a_{i,j}^{t-z}$ is the contagion coefficient from $y_j$ to $y_i$ at lag $z$. A contagion effect from $y_j$ to $y_i$ at lag $z$ can be inferred if $a_{i,j}^{t-z}$ is statistically significant with a confidence level of at least 90$\%$. A positive coefficient means that an increase in one time series is associated with an increase in the other time series after a certain time and vice versa, a negative coefficient means that an increase in one time series is associated with a decrease in the other time series after a certain time and vice versa; $\epsilon_i$ is an error term for the corresponding equation of time series $y_i$.

\subsection{An Early Experiment: Herding Behavior throughout the Entire Period}
We start this study by performing all experiments on the whole examined period without portioning it out. Therefore, based on our research questions, herding behavior detection for the entire period is applied to each type of investment vehicle(i.e. cryptocurrencies, stocks and US ETFs) followed by a detection on the combination of these three types of investment vehicle at a community level.

Table \ref{tab:0} provides the estimation of $\beta_2$ from Eq \ref{eq:4} and $\gamma_2, \gamma_4$ from Eq \ref{eq:5} for each type of investment vehicle. All of them reveal signs of herding behavior over the considered timeline from April 2019 until May 2023 as the coefficient $\beta_2$ is statistically negative in all cases, even though their herding magnitudes are different. When distinguishing the market condition into \textit{up} and \textit{down}, herding is pronounced in both conditions for stocks and US ETFs, however, only the \textit{down} condition in the cryptocurrency market exhibits this phenomenon.

\begin{table}[!h]
\caption{ \bf Herding detection results for each type of investment vehicle over the entire period (01/04/2019 - 03/05/2023). Herding is pronounced when the corresponding coefficient is significantly negative. We consider three levels of confidence 1$\%$ (\textcolor{red}{*}), 5$\%$ (\textcolor{red}{**}) and 10$\%$ (\textcolor{red}{***}) to be significant}
\newcolumntype{C}{>{\centering\arraybackslash}X}
\label{tab:0}
\begin{tabularx}{\textwidth}{CCCC}
\hline
\textbf{Coefficient} & \textbf{Crypto} & \textbf{US ETF} & \textbf{Stock}  \\ \hline
$\beta_2$     & -1.751\textcolor{red}{*}  & -0.859\textcolor{red}{*}  & -3.023\textcolor{red}{*}  \\ 
\begin{tabular}[c]{@{}c@{}}$\gamma_2$\\ (\textit{up} market)\end{tabular}   & 3.465\textcolor{red}{*}   & -1.354\textcolor{red}{*}  & -3.100\textcolor{red}{*}  \\ 
\begin{tabular}[c]{@{}c@{}}$\gamma_4$\\ (\textit{down} market)\end{tabular} & -2.258\textcolor{red}{*}  & -0.854\textcolor{red}{*}  & -2.766\textcolor{red}{*}  \\ 
\hline
\multicolumn{4}{l}{The parameter $\beta_2$ is estimated from the equation:}\\ \multicolumn{4}{l}{$CSAD_t = \alpha + \beta_1|r_{m,t}| + \beta_2 r_{m,t}^2 + \epsilon_t$.}\\
\multicolumn{4}{l}{The parameters $\gamma_2$ and $\gamma_4$ are estimated from the equation:}\\ \multicolumn{4}{l}{$CSAD_t = \alpha + \gamma_1 D^{up} |r_{m,t}| + \gamma_2 D^{up} r_{m,t}^2 + \gamma_3 D^{down} |r_{m,t}| + \gamma_4 D^{down} r_{m,t}^2 + \zeta_t$.} 
\end{tabularx}
\end{table}

\begin{figure}[!h]
\centering
\includegraphics[width = \textwidth]{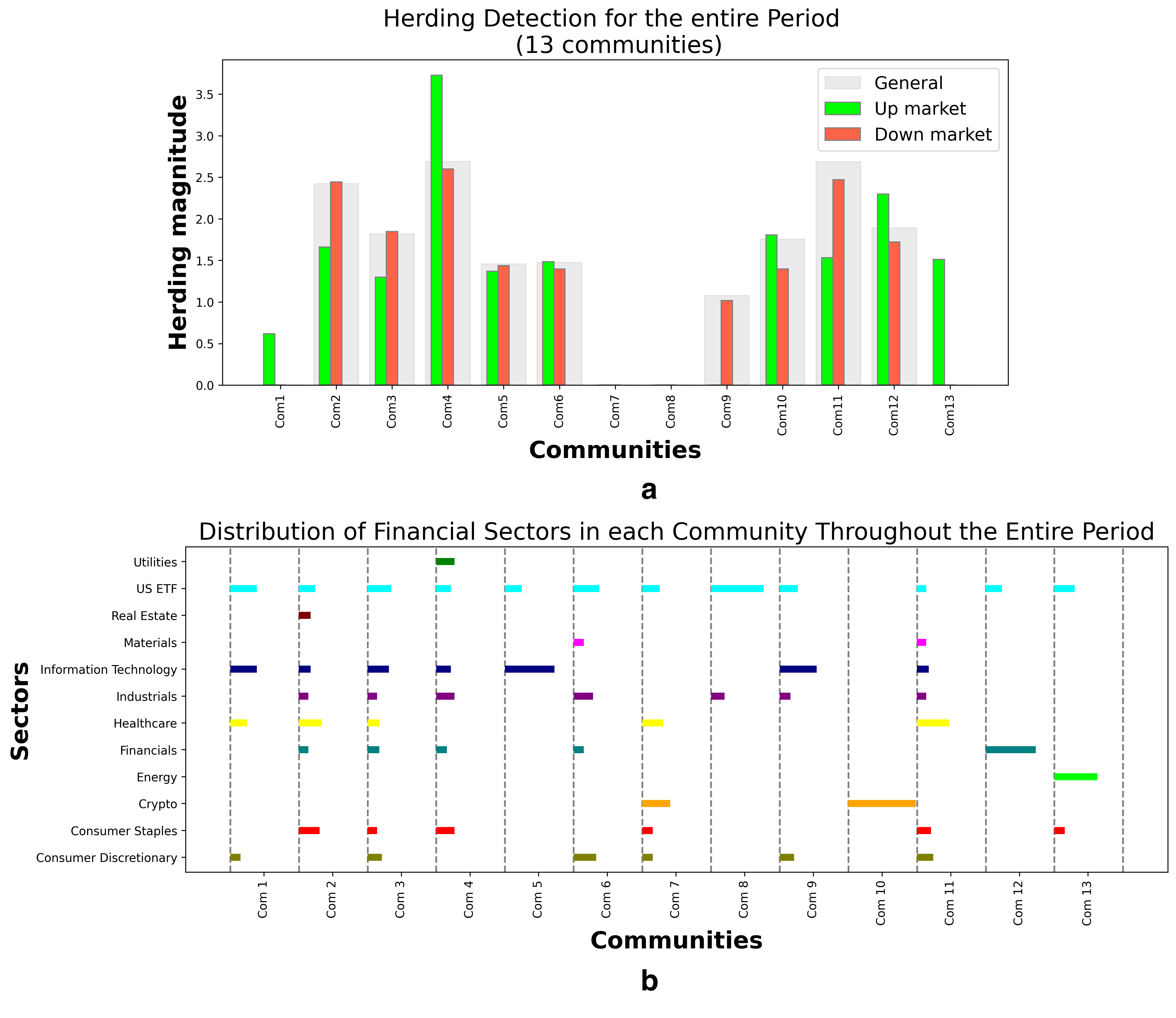}
\caption{\bf Herding detection on each community (a) and its sector distribution (b) over the entire period from 01/04/2019 until 03/05/2023. More information on each sub-figure can be seen at \nameref{SI_Text2}. \label{Fig1}}
\end{figure}

Fig \ref{Fig1} illustrates herding detection results and sector distribution of each community (as a result of applying the Louvain community detection algorithm) in the combination of all three types of assets over the entire period. Specifically, the number of communities found by the Louvain algorithm and the proportion of each sector in each community is illustrated in the lower sub-figure (Fig \ref{Fig1}b). For each community, the dominance of each financial sector within this community is indicated by the length of its corresponding horizontal bar. Sectors are listed on the y-axis, each represented by a unique color, the longer the bar of a corresponding sector, the greater the sector’s percentage within that community. In the meantime, the upper sub-figure (Fig \ref{Fig1}a) reveals the corresponding herding detection results for each community in the lower sub-figure (Fig \ref{Fig1}b). To visualize these herding coefficients, we use their absolute values and refer to them as \textit{herding magnitudes}, if there is no evidence of herding, the herding magnitude is equal to 0. In this regard,  the grey bar means the herding based on CSAD model (Eq~\ref{eq:4}) is significant, the green bar means the herding based on CSAD for the \textit{up} market condition (Eq \ref{eq:5}) is significant, the red bar means the herding based on CSAD for \textit{down} market condition (Eq \ref{eq:5}) is significant. The higher the bar, the more pronounced the herding. The assets in each community are listed in \nameref{S3_Tab}. Overall, it is shown that performing herding detection on the whole period tends to finish with a 'Yes' result. However, it is almost impossible to conclude anything given this result since the examined period is long, leading to a lack of detail. Furthermore, all types of investment vehicle exhibit herding behavior so no unique patterns can be found. This ambiguity encourages us to split the original period into sub-periods so that the herding behavior can be examined on a finer time interval, which enhances the detail level of the experiment. The information for this timeline division is proposed in the next section. 

\subsection{Timeline Division}
As mentioned earlier, our dataset covers a 4-year period from 01/04/2019 to 03/05/2023. Throughout these years, the financial markets experienced several pressing events, including the US-China trade war~\cite{che23,bai23,ben23}, the  Covid-19 pandemic~\cite{dro20,ach21,li22}, and the Ukraine-Russia conflict~\cite{conlon24,der23,das23,kum23}.  Since different events are characterized by different behavioral trends/reactions of investors, we split the given period into smaller intervals based on the events that occurred. Following this, herding detection-related experiments in this study are run on each sub-period. This practice helps us track the changes in herding behavior over time as well as recognize common features appearing during a particular event.

Notably, after Covid-19's first wave, the financial markets began to enter the recovery time and most assets in both traditional and cryptocurrency markets have witnessed a surge in prices since then. This was a period when most assets reached their all-time highest values. Since 2022, the markets have calmed down gradually, marking the end of the bullish time for most assets~\cite{wor21}. Thus, we also consider this bull market time in our timeline division. 

To this end, we partition the original timeline into 7 sub-periods as shown in Table \ref{tab:3}. We provide some explanations for this division of sub-periods in this table. Firstly, we choose the date 30/06/2020 as a cutoff for the Covid-19 pandemic sub-period because the financial markets have recovered to their Pre-Covid-19 performance and the general public attention towards the pandemic has been lessened significantly in July 2020~\cite{pha22}. Since then, both cryptocurrency and traditional markets continued surging in prices until the end of 2021. In other words, the financial markets entered bull market conditions after overcoming the severe hit by the global economic crisis in March 2020. Ideally, the period from 01/07/2020 to 23/02/2023 (before the Ukraine-Russia conflict onset) can be called a \textit{bull market time}, but this period is too long. Therefore, we divide it into 3 sub-periods. Lastly, the remaining time coincides with the Ukraine-Russia conflict event. However, this period is also too long compared to the previous ones. Therefore, we divide it into 2 sub-periods.

\begin{table}[h!]
\centering
\caption{Timeline division in this study. The original period (01/04/2019 - 03/05/2023) is segmented into 7 sub-periods based on key historical events and market conditions}
\label{tab:3}
\begin{tabular}{ccc}
\hline
\textbf{Sub-period} & \textbf{Name}                      & \textbf{Time Interval}            \\ \hline
1          & Pre-Covid-19              & 01/04/2019 to 31/12/2019 \\
2          & Covid-19 Pandemic         & 01/01/2020 to 30/06/2020 \\
3          & Bull Time 1               & 01/07/2020 to 31/01/2021 \\
4          & Bull Time 2               & 01/02/2021 to 31/08/2021 \\
5          & Bull Time 3               & 01/09/2021 to 23/02/2022 \\
6          & Ukraine-Russia Conflict 1 & 24/02/2022 to 30/09/2022 \\
7          & Ukraine-Russia Conflict 2 & 01/10/2022 to 03/05/2023 \\
\hline
\end{tabular}
\end{table}

\section{Experimental Results and Discussion} \label{sec:4}
This section investigates the herding behavior in cryptocurrencies, US ETFs and stocks, across different sub-periods. The results are then synthesized to explore insights about this phenomenon. Our experiments are structured into two main tasks according to the defined research questions: Firstly, we implement herding detection in each separate type of investment vehicle (RQ1). Secondly, we combine all three aforementioned types of investment vehicle and perform herding detection on the combination (RQ2). This idea comes from the hypothesis that the view on each type of investment vehicle differs from a composition of multiple different ones. In particular, knowing the herding patterns of a specific market is important for understanding its unique features, major events that the market has experienced, and which kind of information/news impacts the market, etc. On the other hand, observing different types of investment vehicle at the same time is critical for understanding the connections between them, which can highlight the way investors diversify their portfolios or the impact of each separate type of investment vehicle on their combination. Such insights cannot be obtained by focusing only on a single market.

\subsection{Herding Detection on Each Type of Investment Vehicle}

In this part, we tackle the first research question by uncovering herding signals in each type of investment vehicle across the seven aforementioned sub-periods. Subsequently, answering various sub-questions such as: Which events cause the phenomenon? How strong is it? How are investors' reactions linked to the herding? How resilient is the herding? Then, a comparison between the three types of investment vehicle is conducted.

Table \ref{tab:4.1} provides the estimation of $\beta_2$ from the CSAD model (without distinguishing \textit{up} and \textit{down} market conditions) while Table \ref{tab:4.2} provides the estimation of  $\gamma_2$, and $\gamma_4$ from CSAD for the \textit{up} market condition and CSAD for the \textit{down} market condition, respectively, for each type of investment vehicle. Only values with the star symbol are statistically significant. Herd behavior is revealed at a period only when its corresponding coefficient (either $\beta_2$,  $\gamma_2$ or $\gamma_4$) is statistically significantly negative. The more negative the value, the more pronounced the herding.

\begin{table}[!h]
\centering
\caption{ \bf Herding detection results for each type of investment vehicle, without distinguishing \textit{up} and \textit{down} market conditions. Herding exists in a period if the corresponding coefficient $\beta_2$ is significantly negative. We consider three levels of confidence (1$\%$ (\textcolor{red}{*}), 5$\%$ (\textcolor{red}{**})) and 10$\%$ (\textcolor{red}{***}) to be significant}
\label{tab:4.1}
\begin{tabular}{>{\centering\arraybackslash}p{4cm}ccc}
\hline
 &
  \multicolumn{3}{c}{\textbf{CSAD ($\beta_2$)}}\\
  \hline
\textbf{Periods} &
  \textbf{Crypto} &
  \textbf{US ETF} &
  \textbf{Stock} \\

  \hline
Pre-Covid-19    & 6.3223\textcolor{red}{*}  & -4.5498\textcolor{red}{*}     & -9.1788\textcolor{red}{*}  \\
Covid-19 Pandemic & -4.7204\textcolor{red}{*} & -0.9952\textcolor{red}{*}      & -2.7261\textcolor{red}{*} \\
Bull Time 1   & -6.7795\textcolor{red}{*} & 8.1034\textcolor{red}{*}     & 9.2020\textcolor{red}{*}   \\
Bull Time 2      & -3.7911\textcolor{red}{*} & 38.3803\textcolor{red}{*}    & -10.1002\textcolor{red}{**}      \\
Bull Time 3 & 1.0469\textcolor{red}{*}  & 1.5452 & 3.8338     \\
U-R Conflict 1 & 0.4652\textcolor{red}{**} & 1.6437\textcolor{red}{**} & 5.9794\textcolor{red}{*}  \\ 
U-R Conflict 2 & -9.1943\textcolor{red}{*} & -0.4417 & -0.5448 \\ 
\hline
\multicolumn{3}{l}{The parameter $\beta_2$ is estimated from the equation $CSAD_t = \alpha + \beta_1|r_{m,t}| + \beta_2 r_{m,t}^2 + \epsilon_t$.}  
\end{tabular}%
\end{table}

\begin{table}[!h]
\centering
\caption{ \bf Herding detection results for each type of investment vehicle when considering \textit{up} and \textit{down} market conditions separately. Herding is pronounced during the positive (negative) market returns if the corresponding coefficient $\gamma_2$ ($\gamma_4$) is significantly negative. We consider three levels of confidence 1$\%$ (\textcolor{red}{*}), 5$\%$ (\textcolor{red}{**}) and 10$\%$ (\textcolor{red}{***}) to be significant}
\label{tab:4.2}
\begin{tabular}{>{\centering\arraybackslash}p{4cm}cccccc}
\hline
 &
  \multicolumn{3}{c}{\textbf{\textit{Up} Market Condition ($\gamma_2$)}} &
  \multicolumn{3}{c}{\textbf{\textit{Down} Market Condition ($\gamma_4$)}} \\
  \hline
\textbf{Periods} &
  \textbf{Crypto} &
  \textbf{US ETF} &
  \textbf{Stock} &
  \textbf{Crypto} &
  \textbf{US ETF} &
  \textbf{Stock} \\
  \hline
Pre-Covid-19    &  9.1014\textcolor{red}{*}  & -2.0638    & -1.6660     & 4.5191\textcolor{red}{*}  & -5.0853\textcolor{red}{*}     & -10.0953\textcolor{red}{*} \\
Covid-19 Pandemic        &  -4.3471\textcolor{red}{*} & -1.2972\textcolor{red}{*}    & -3.2433\textcolor{red}{*}  & -4.6894\textcolor{red}{*} & -0.9982\textcolor{red}{*}     & -2.5910\textcolor{red}{*} \\
Bull Time 1   &  -6.4200\textcolor{red}{*} & 8.9108\textcolor{red}{*}    & 7.0201\textcolor{red}{*}   & -6.1360\textcolor{red}{*} & -0.5774     & 6.5425\textcolor{red}{**}  \\
Bull Market 2      & 0.4969  & 10.9087\textcolor{red}{*} & -3.3375 & -3.9754\textcolor{red}{*} & 38.9335\textcolor{red}{*}   & -14.5454\textcolor{red}{*}    \\
Bull Time 3 &  2.2923\textcolor{red}{*}  & -5.5150\textcolor{red}{**}     & -0.5228    & 1.0751\textcolor{red}{*}  & 3.5424\textcolor{red}{*} & 6.8038\textcolor{red}{*}    \\
U-R Conflict 1 & 0.0358 & 7.3115\textcolor{red}{*} & 15.5605\textcolor{red}{*} & 0.6319\textcolor{red}{*} & 0.6063 & 3.7276\textcolor{red}{***} \\
U-R Conflict 2 & -11.9781\textcolor{red}{*} & -0.2959 & -0.7646 & -7.8995\textcolor{red}{*} & -0.5792 & -0.3428 \\
\hline
\multicolumn{6}{l}{The parameters $\gamma_2$ and $\gamma_4$ are estimated from the equation}\\ \multicolumn{6}{l}{$CSAD_t = \alpha + \gamma_1 D^{up} |r_{m,t}| + \gamma_2 D^{up} r_{m,t}^2 + \gamma_3 D^{down} |r_{m,t}| + \gamma_4 D^{down} r_{m,t}^2 + \zeta_t$.} 
\end{tabular}%
\end{table}

Starting with the cryptocurrency market, the herding coefficient $\beta_2$ reveals herding signals in four sub-periods, including during the Covid-19 Pandemic, the first and second Bull Time, and the second Ukraine-Russia Conflict. Surprisingly, the strongest herding signals are found in the first Bull Time and the second Ukraine-Russia Conflict sub-periods while the peak of the Covid-19 pandemic, which we expected to show the strongest herding behavior, only comes in third place. On the other hand, the Pre-Covid-19, the third Bull Time and the first Ukraine-Russia Conflict sub-periods disclose an absence of herding behavior. The results remain mostly unchanged after separating the market into \textit{up} and \textit{down} conditions. Specifically, the coefficients $\gamma_2$ and $\gamma_4$ are only statistically significantly negative during the sub-periods when $\beta_2$ is negative, indicating that either the \textit{up} or \textit{down} market condition or both conditions reveal the evidence of herding in these stages. However, while herding is pronounced in both \textit{up} and \textit{down} market conditions in most cases, only the downturn times in the second sub-period where Bull Market conditions prevail have the appearance of herding. These results sufficiently capture the prevailing sentiment among cryptocurrency investors. Specifically, they seem to be influenced by market fluctuations and breaking news\cite{da19}, leading to similar investment decisions, i.e. selling off their investments when the market experiences a downturn or bad news circulates on the internet to avoid a loss while purchasing new investments during bullish phases or when positive news about cryptocurrencies spreads among the general public due to the Fear of Missing Out (FOMO) on an increase in the price of the coins\cite{jal20}. The high degree of sensitivity of investors in this digital market may be attributed to the dominance of irrational (na\"ive) participants, as confirmed in different studies \cite{alm23}, these people participate in the cryptocurrency market primarily because of its great potential for huge profit despite their lack of investment experience. Therefore, they tend to ignore their own analysis and information by following the decisions made by the public due to their lack of financial knowledge \cite{kau23}. A greater herding magnitude during the first Bull Time and the second Ukraine-Russia Conflict sub-periods compared to the peak of the pandemic (Covid-19 Pandemic sub-period) can be attributed to the cryptocurrency market's tendency to be more influenced by positive circumstances than negative ones \cite{kaz19, aha21}. Specifically, during the first Bull Time, the market started to recover from the decline caused by the pandemic, experiencing a price surge since then. An upward trend in prices along with positive word of mouth, as well as financial news, triggered the positive sentiment of investors together with the FOMO phenomenon that eventually caused a purchase herding in the market \cite{kau23,wan23,wak22}. When it comes to the second Bull Time, although the herding is evident, it appears to be weaker than other sub-periods. This might be because many investors had already purchased the coins at the earlier stage and were more likely to hold onto those coins during this time because of the positive expectations of an upward trend in the future. Thus, they barely purchased at this time, instead, the herding might come from them executing a sell action. Indeed, we notice that there was a downturn phase in this sub-period when the market significantly shrank.  Hence, it is possible that a portion of investors began selling off their investments as they either felt panicked by the price drop or were motivated to take profits, resulting in a period of herding \cite{jal20,you22,bal20,jal20}. Our result when examining the cryptocurrency market in \textit{up} and \textit{down} conditions also shows that only $\gamma_4$ is negative during the second Bull Time, meaning that herding behavior is much more intense during the downturn time compared to the upturn one, which reinforces our hypothesis. Moreover, the same circumstance is also seen in the second Ukraine-Russia Conflict sub-period. During this time, financial markets experienced an upward trend and cryptocurrencies are no exception to this. Specifically, herding behavior in the cryptocurrency market is found in both \textit{up} and \textit{down} market conditions with the herding being more intense during times of positive market returns. Overall, results of herding detection among cryptocurrencies market are in line with existing literature which also demonstrates evidence of herding during the first outbreak of the pandemic/ 2020 global economic crisis and the periods of bullish markets \cite{man22,wak22,wan23,pap21}.

The herding behavior observed in the stock and US ETF markets displays similar patterns most of the time, with the first two sub-periods, namely Pre-Covid-19 and Covid-19 Pandemic, having signs of herding behavior while most remaining sub-periods show no evidence of herding.These first two sub-periods cover critical events that significantly impacted traditional markets, i.e. the US-China trade war during the year 2019 and the severe hit of the contagious health disease Covid-19 on the global economy and financial markets in the first half of 2020. Notably, in the Pre-Covid-19 period (Tables \ref{tab:4.1} and \ref{tab:4.2}), herding behavior in both stocks and US ETFs was significantly stronger during times of negative market returns than positive ones. This greater herding magnitude can be attributed to the tension of the US-China trade war throughout 2019, affecting not only the US and Chinese economies but also other countries and continents~\cite{ben23,bai23}. Additionally, investors were continuously exposed to negative news about the trade war throughout the year~\cite{lew19}. Consequently, these circumstances led to a herding phenomenon among investors, especially during periods of declining asset prices. In fact, it has been shown that there was a selling wave of assets during this year~\cite{par19}. The only exception was that stocks also exhibited a signal of herding behavior during the second Bull Time while US ETFs revealed evidence of this behavior during the third Bull Time. This herding behavior can be attributed to the continual surge in prices of the assets during these sub-periods~\cite{wor21}, which caused similar purchasing and selling actions among investors with the aim of making the most profits. 

The herding in the stock and US ETF markets delivers two interesting results. Firstly, herding is more pronounced during the Pre-Covid-19 compared to the Covid-19 Pandemic sub-period. Therefore, although the pandemic led to greater losses and fluctuations in traditional financial markets, it is noteworthy that investors exhibited more intense herding behavior before the pandemic, particularly during the turbulent time caused by the US-China trade war. One possible reason for this is the constant fluctuations of markets throughout 2019 when the trade war became more and more intense. In particular, different tariffs and quotas were imposed between these two countries throughout this year, directly impacting financial markets worldwide. As a result, the constant release of negative and positive news gave rise to frequent market movements. Thus, investors actively reacted to market situations, causing herding behavior on a regular basis. The impact of the Covid-19 pandemic was truly severe and damaging within the second quarter of 2020 but this was followed by a gradual recovery. Hence, investors panicked during the peak of the pandemic but things quickly got back to normal, and as they started to make objective investment decisions again, herding then lessened. Another possible reason is that investors in traditional markets are more influenced by macroeconomic factors as well as policy uncertainty. Specifically, US interest rates increased in response to rising inflation. Furthermore, the unemployment rate increased due to the trade war and GDP in 2019 also fell to its lowest since the 2008 crisis~\cite{mal20,cho19,ahn24}. These macroeconomic factors had the potential to scare traditional market traders. 

Secondly, the magnitude of herding behavior in the market composed of US ETFs is generally weaker than in the stock and cryptocurrency markets, as indicated by the lower herding coefficients ($\beta_2, \gamma_2, \gamma_4$) for US ETFs compared to those for stocks and cryptocurrencies across most sub-periods. This distinction is possibly driven by the stylized facts of US ETFs: the US market itself has a well-established economic environment and a high proportion of rational investors \cite{chi10,yan23}. As a result, since investors are more inclined to rely on their own analysis and expectations to make investment decisions rather than mimicking other investors, the herding behavior is therefore reduced. This large number of rational investors may be a crucial factor in limiting herding behavior, since the herding frequency in stocks and US ETFs is lower than in cryptocurrencies, as evidenced by Tables \ref{tab:4.1} and \ref{tab:4.2}. On the other hand, another stylized fact about ETFs is their low volatility (i.e. low risk) compared to other types of investment vehicle, which limits the potential for short-term profits. Therefore, investors tend to herd in other - more risky - types of investment vehicle to gain high profits while ETFs are preferable for long-term investments~\cite{mei23}. Remarkably, also in US ETFs, Tables \ref{tab:4.1} and \ref{tab:4.2} display an intense dispersion in price movements between these assets during the second Bull Time sub-period, with $\beta_2$ exceeding 38.0 and $\gamma_4$ nearing 39.0. This indicates a substantial divergence in investment decisions among investors in US ETFs during this time.

In summary, we found that investors in all three types of investment vehicle showed evidence of herding during the Covid-19 Pandemic sub-period. This is reasonable since the serious pandemic directly affected the global economy and all financial markets. However, most remaining sub-periods reveal different herding patterns between the cryptocurrency market and traditional markets. In particular, herding behavior was found only in stocks and US ETFs throughout 2019, while this phenomenon disappeared in the cryptocurrency market. Indeed, herding during this time appeared to be caused by the trade war between China and the US which was relevant to macroeconomic factors and directly influenced various companies in different sectors. Thus, traditional markets were undoubtedly affected the most. On the other hand, only the cryptocurrency market displayed a herding signal during the first Bull Time and second Ukraine-Russia Conflict sub-periods. This can be explained by the fact that cryptocurrency investors are often swayed more by positive news and positive market returns than negative ones~\cite{poy18}. On the other hand, traditional markets are typically more responsive to distressed market conditions than booms~\cite{duy21,aha21}. In particular, when traditional markets experience an upward trend, professional investors are found to account for the majority portion~\cite{ahn24}. Thus, they tend to make investment decisions by themselves which prevents them from performing as a herd. For the cryptocurrency market, however, its investors are more likely to be na\"ive and make irrational decisions, based on word of mouth as well as existing news on social media~\cite{ngu23,alm23}. Thus, positive news surrounding cryptocurrencies can lure inexperienced investors with the promise of substantial profits.

\subsection{Herding Detection on a Combination of Cryptocurrencies, Stocks and US ETFs}

In reality, investors tend to diversify their portfolios by investing in different types of assets~\cite{con24}. Taking inspiration from this practice, this section combines all three types of investment vehicle used in this study to gain a more complete and realistic picture of herding behavior in financial markets, which aims to answer our second research question. However, it is unwise to conclude anything about the herding behavior of this large dataset whose assets belong to three different types of investment vehicle. Besides,  there is no established approach to partition assets into distinct subsets for herding detection currently. To address this,  we come up with the idea of using community detection to split the assets into smaller communities depending on their similarities. Subsequently, we apply herding detection techniques to each community. This practice is grounded in the belief that assets with common price movements are more likely to show herding behavior. Conversely, assets with different price movements tend to be used as safe havens against each other rather than being herded~\cite{sha19}. Thus, splitting the assets in this way can preserve the characteristics of portfolio diversification and enhance the clarity as well as granularity of herding signals while mitigating the chance of losing possible herding signals.

Furthermore, we provide information about sector distribution for each community. Specifically, each asset in a community is categorized into either cryptocurrency, or US ETF, or one of the 11 stock sectors discussed in section~\nameref{sec:3}. Subsequently, the proportion of each sector is calculated in each community. Finally, for each community in each sub-period, we can find whether herding behavior exists in those assets and its corresponding sector distribution.

\subsubsection{Community Structures for a Combination of Cryptocurrencies, Stocks and US ETFs}

\begin{figure}[H]
\centering
\includegraphics[width= 9.5cm]{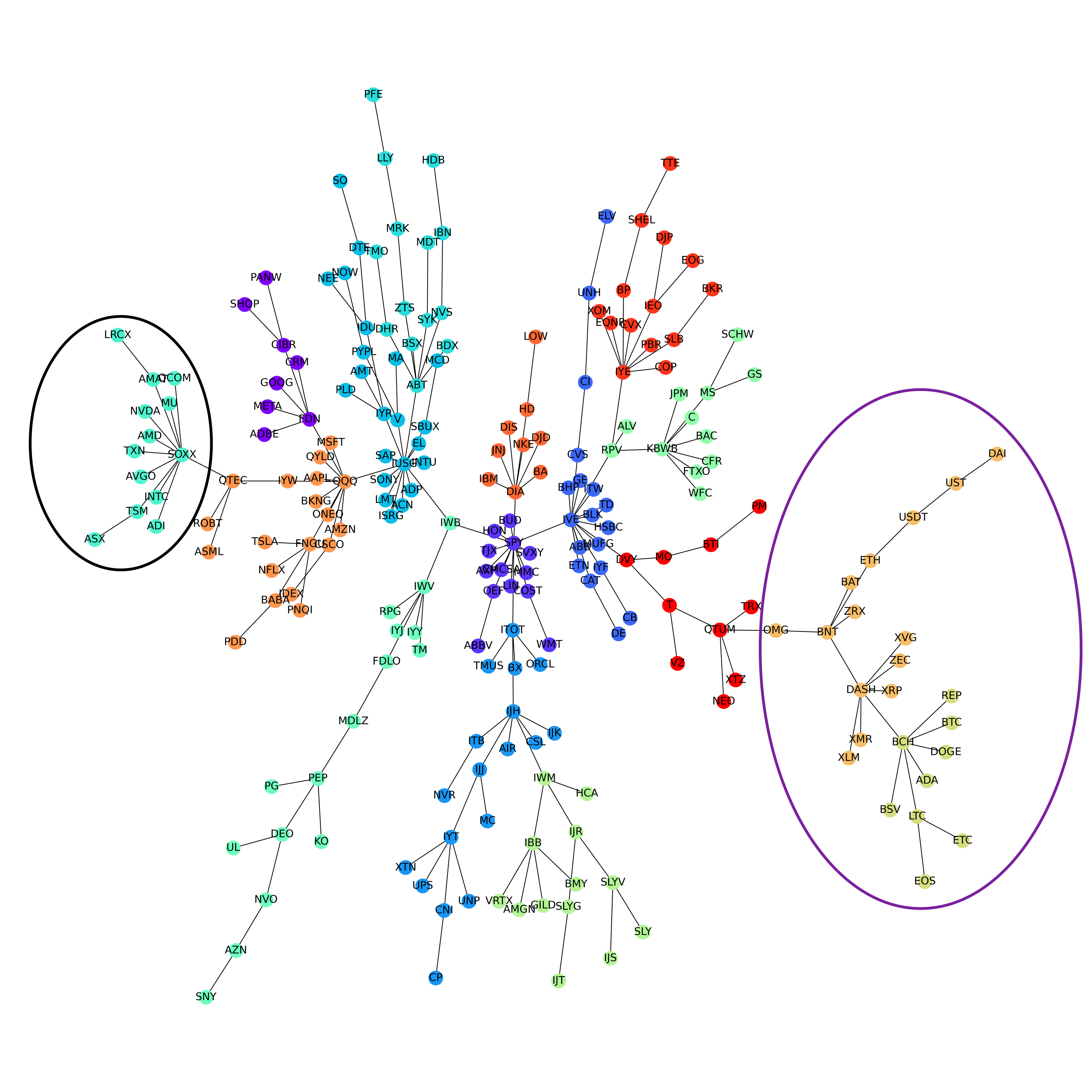}
\caption{ \bf Community structure of cryptocurrencies, stocks and US ETFs during the Pre-Covid-19 period. More information can be seen at \nameref{SI_Text3}}\label{Fig2}
\end{figure}

\begin{figure}[H]
\centering
\includegraphics[width= 9.5cm]{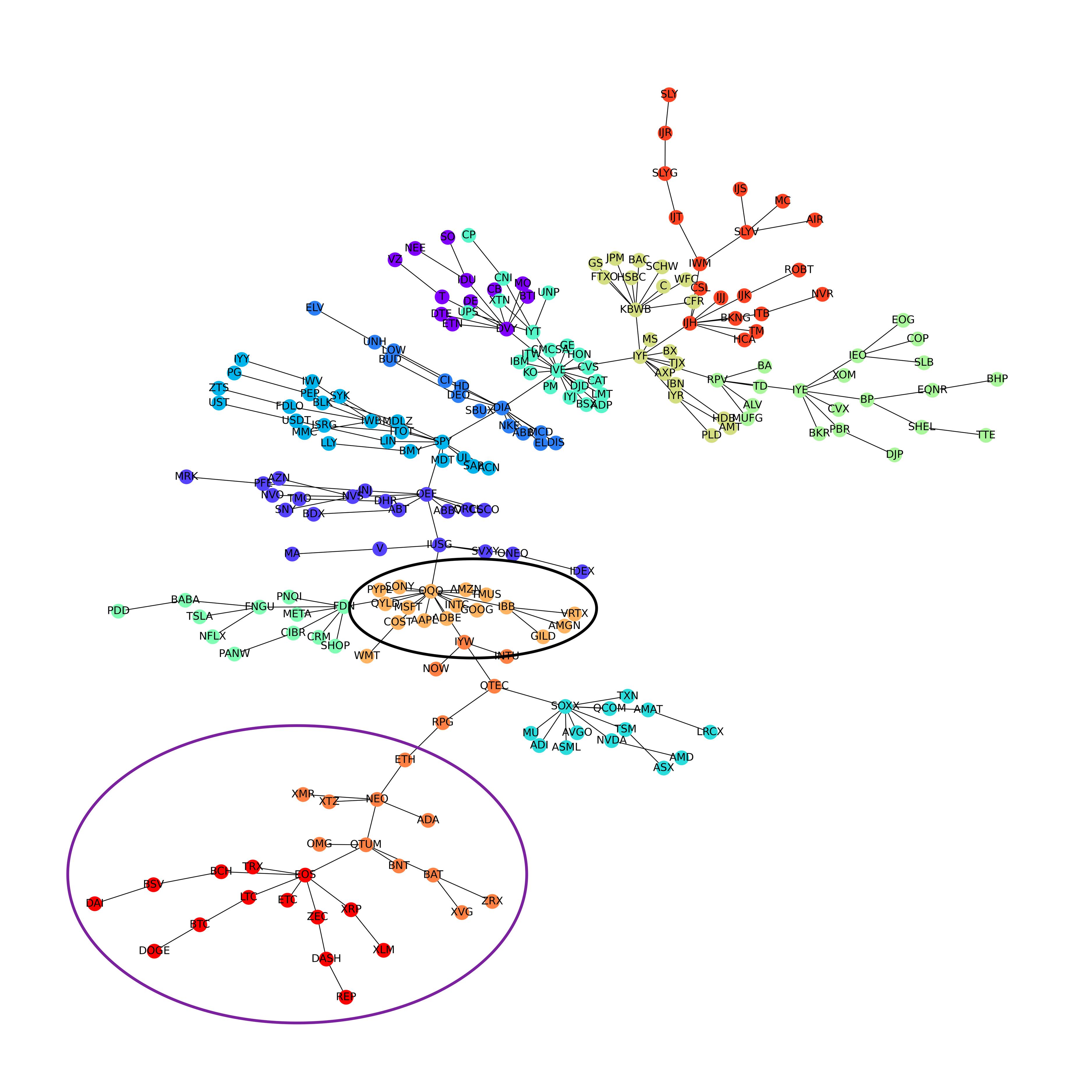}
\caption{\bf Community structure of cryptocurrencies, stocks and US ETFs during the Covid-19 Pandemic period. More information can be seen at \nameref{SI_Text3}}\label{Fig3}
\end{figure}

\begin{figure}[H]
\centering
\includegraphics[width= 9.5cm]{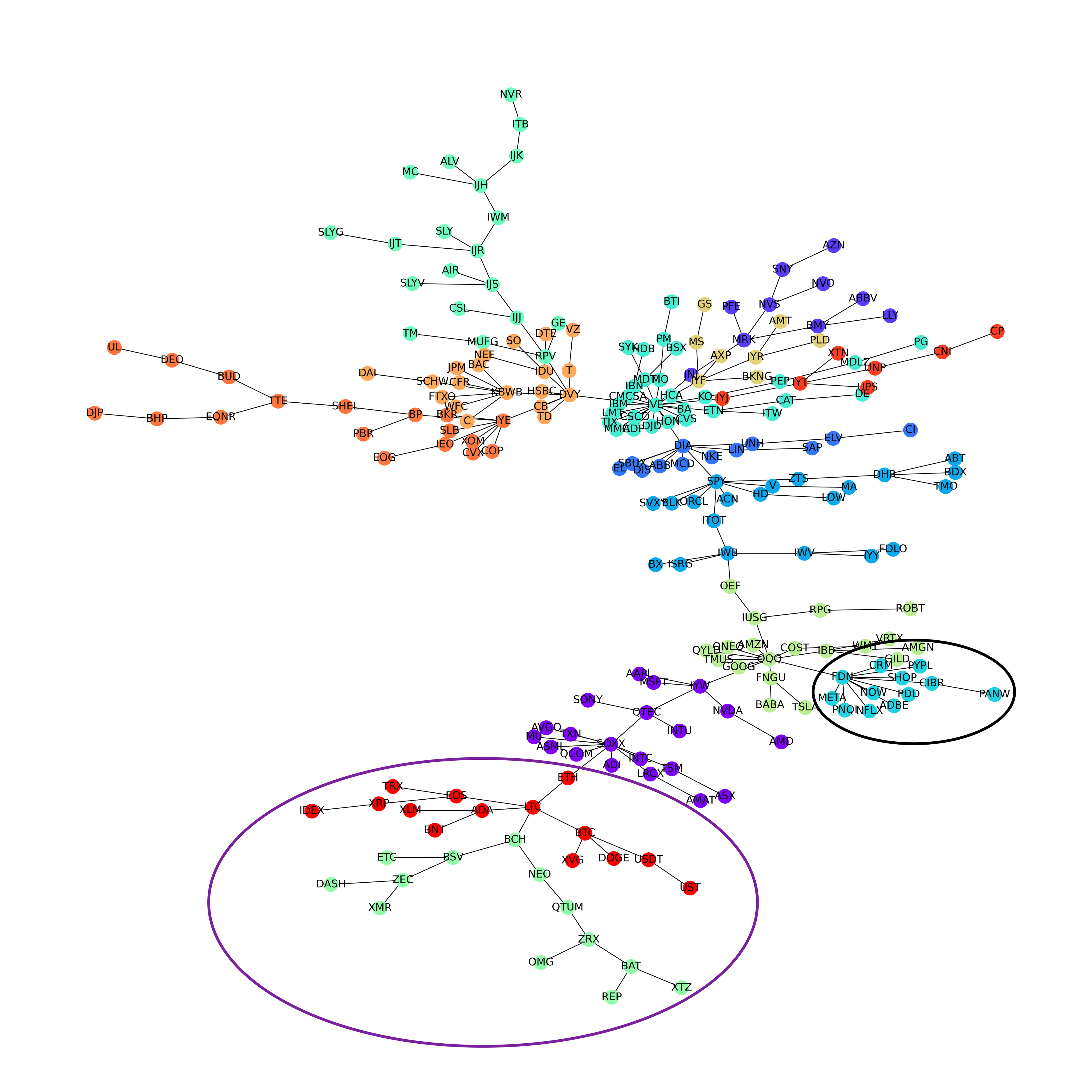}
\caption{\bf Community structure of cryptocurrencies, stocks and US ETFs during the first Bull Time period. More information can be seen at \nameref{SI_Text3}}\label{Fig4}
\end{figure}

\begin{figure}[H]
\centering
\includegraphics[width= 9.5cm]{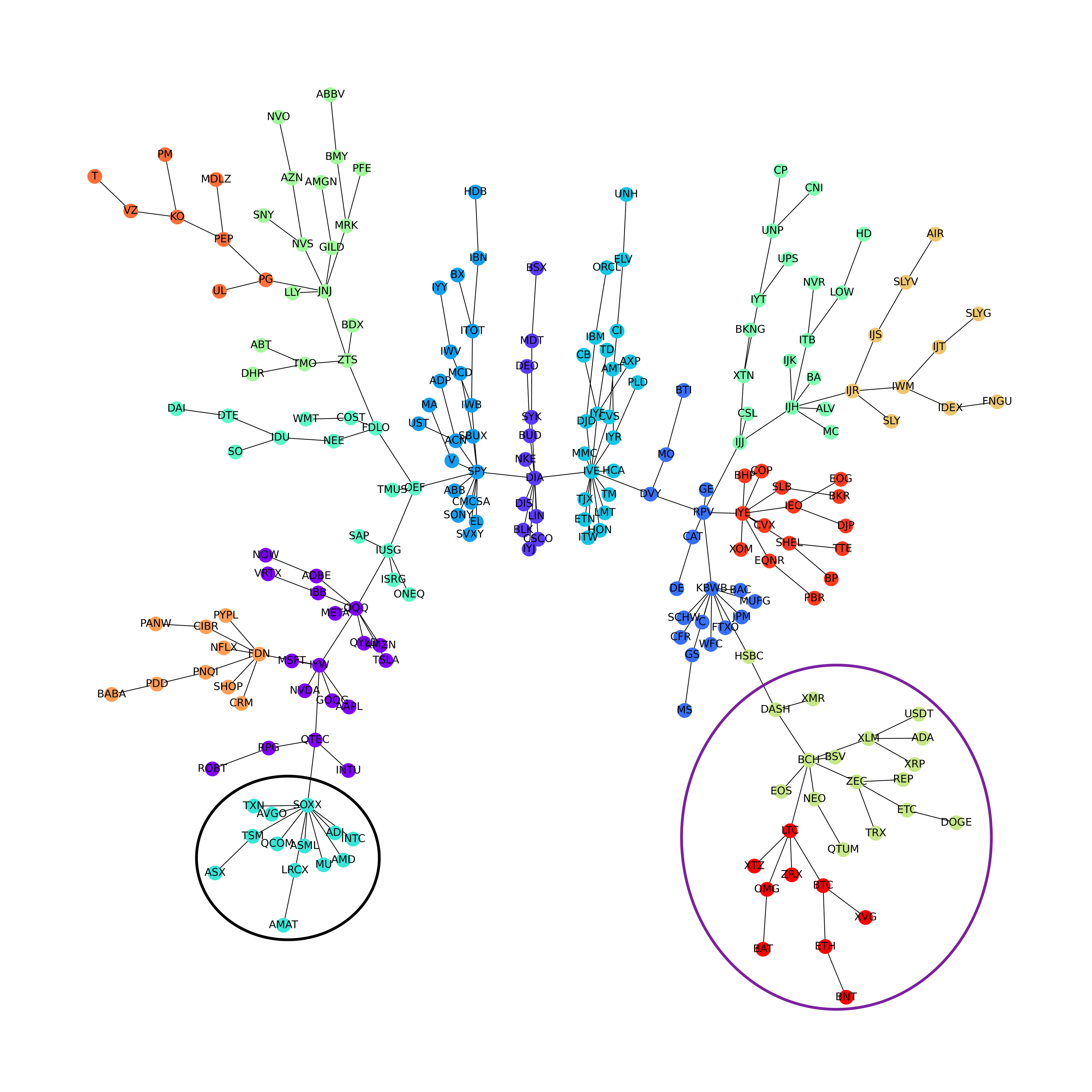}
\caption{\bf Community structure of cryptocurrencies, stocks and US ETFs during the second Bull Time period. More information can be seen at \nameref{SI_Text3}}\label{Fig5}
\end{figure}

\begin{figure}[H]
\centering
\includegraphics[width= 9.5cm]{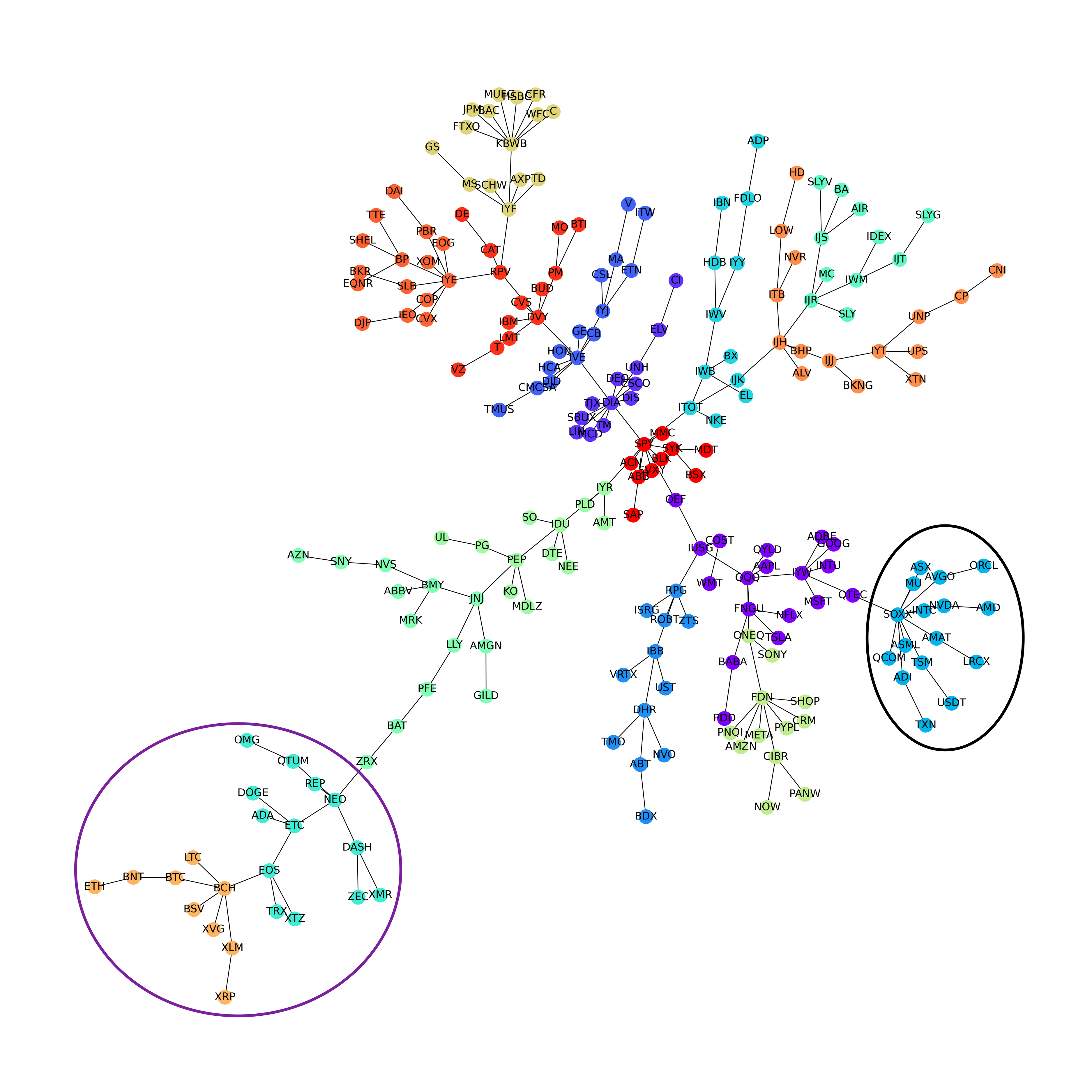}
\caption{\bf Community structure of cryptocurrencies, stocks and US ETFs during the third Bull Time period. More information can be seen at \nameref{SI_Text3}}\label{Fig6}
\end{figure}

\begin{figure}[H]
\centering
\includegraphics[width= 9.5cm]{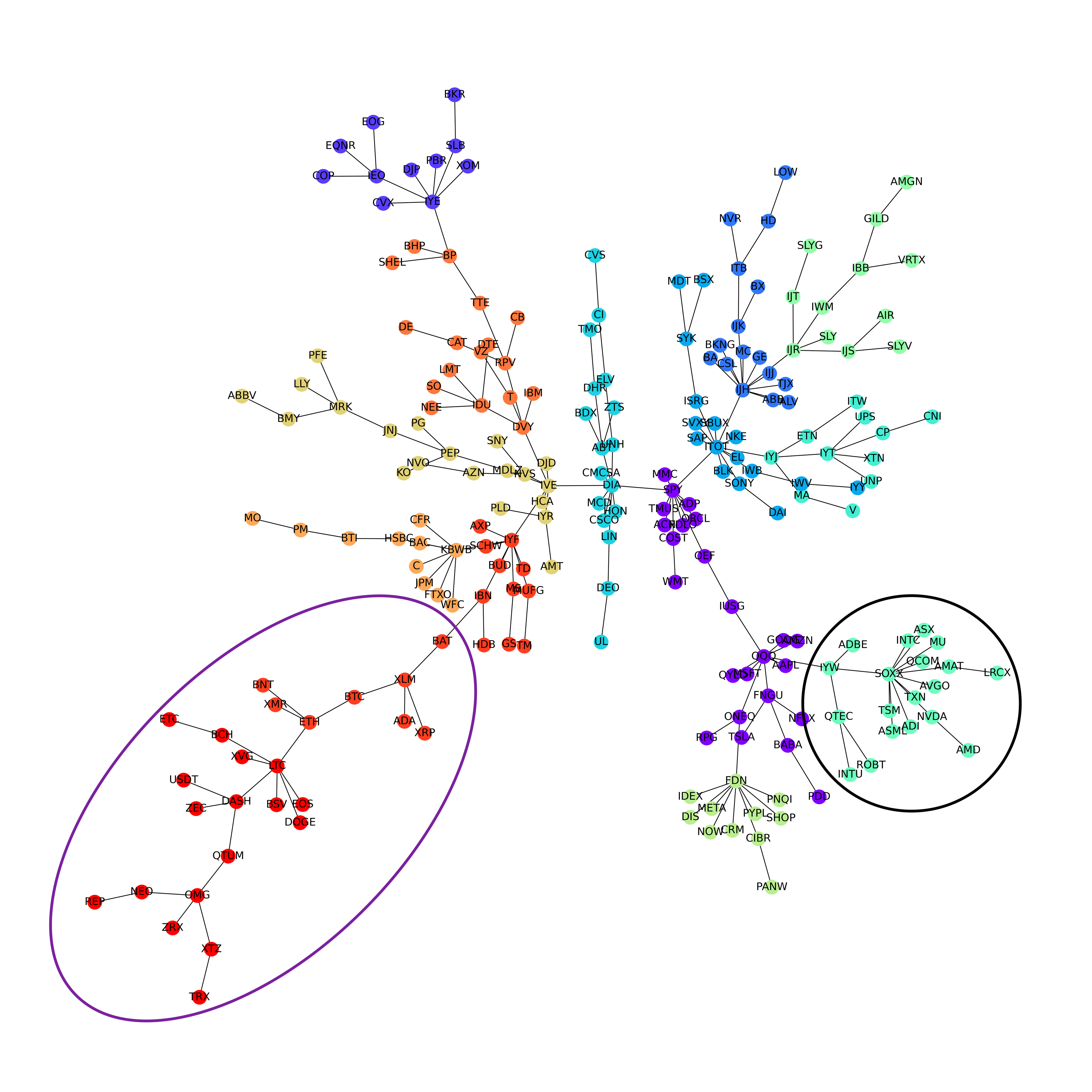}
\caption{\bf Community structure of cryptocurrencies, stocks and US ETFs during the first Ukraine-Russia period. More information can be seen at \nameref{SI_Text3}}\label{Fig7}
\end{figure}

\begin{figure}[H]
\centering
\includegraphics[width= 9.5cm]{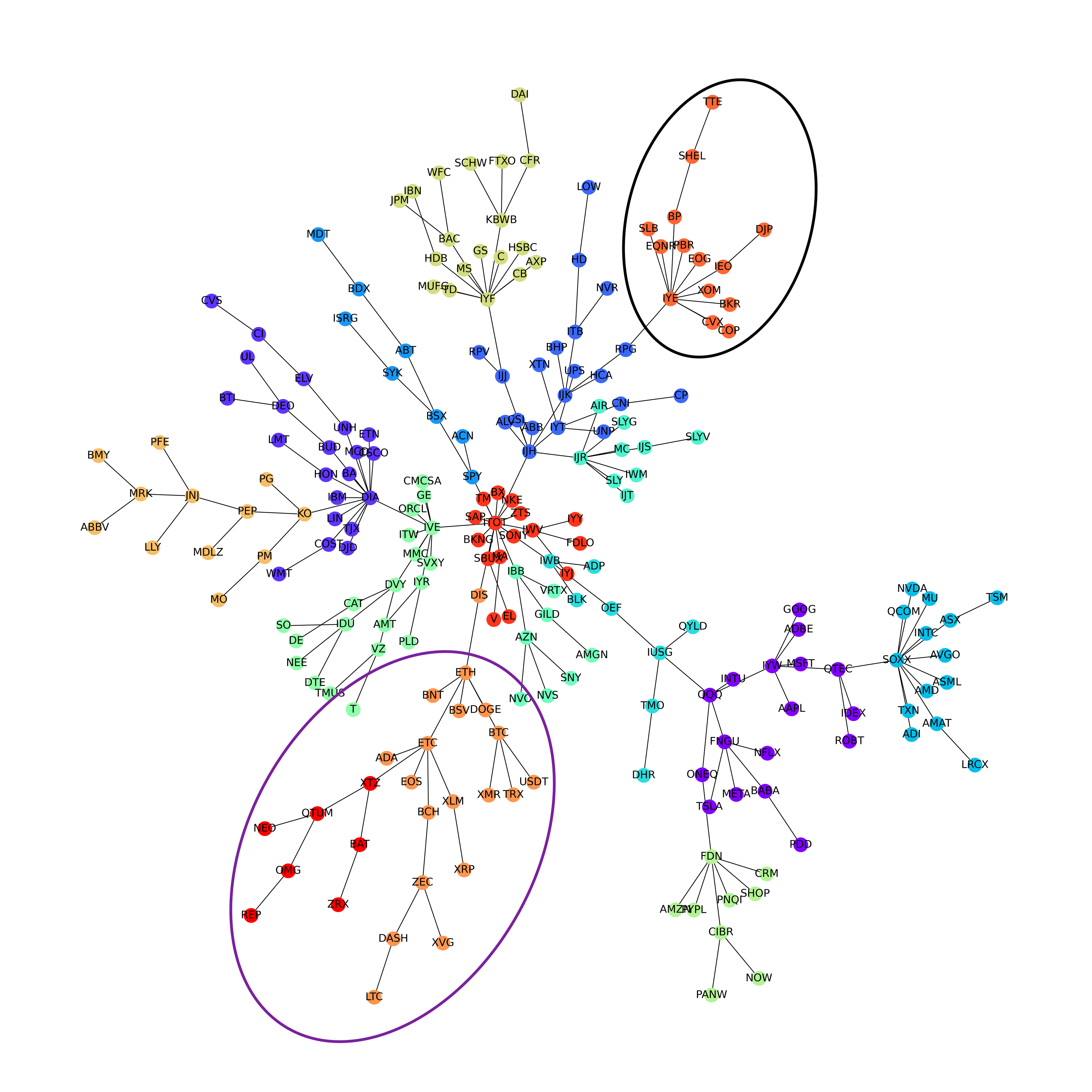}
\caption{\bf Community structure of cryptocurrencies, stocks and US ETFs during the second Ukraine-Russia period. More information can be seen at \nameref{SI_Text3}}\label{Fig8}
\end{figure}

We first introduce the community structure obtained from a combination of cryptocurrencies, stocks and US ETFs in each sub-period, which is displayed in Figs \ref{Fig2}-\ref{Fig8}. Each node in the graph represents an asset, while an edge between two nodes represents their correlation. If there is no edge between two nodes, then their correlation is negligible compared to other surrounding linked nodes. Each color represents a community in which assets with the same color belong to one community, meaning that they behave similarly and their intercorrelations are stronger than those of correlations with other nodes in different communities. The longer the path between two nodes, the more dissimilar they are.

As can be seen in Figs \ref{Fig2}-\ref{Fig8}, the community structures change over time, indicating the variation in correlations among assets over time. Notably, we found some common patterns out of these seven different structures. Firstly, cryptocurrencies form several separate communities but these are next to each other, meaning that there is a strong correlation between cryptocurrencies themselves. Furthermore, there is still a distinction between the cryptocurrency market and traditional markets since most cryptocurrencies do not form communities with stocks and US ETFs, an example of this characteristic is highlighted within a purple-shaped oval in each figure. Secondly, it can be seen that stocks and US ETFs tend to belong to the same community. Interestingly, US ETFs seem to be the centre of a community, linked by stock assets, indicating that they are the most influential nodes in a community, an example of this characteristic is highlighted within a black-shaped oval in each figure. Thirdly, there are some stock sectors in which stocks within the same sector and ETFs that contain these stocks tend to be in the same communities, namely Energy, Healthcare, Information Technology and Financials. On the other hand, this feature does not appear in the remaining sectors. Fourthly, we also notice that S\&P-related ETFs tend to belong to one community throughout the considered sub-periods.

\subsubsection{Herding Behavior at Community Level}

We note that due to a large number of herding results, it is challenging to compare general CSAD (Eq \ref{eq:4}), CSAD for \textit{up} and CSAD for \textit{down} market conditions (Eq \ref{eq:5}) in each community, between different communities and between different sub-periods. Therefore, we do not differentiate the results of general herding, herding during the \textit{up} and herding during the \textit{down} market condition, i.e. we do not go into detail about CSAD, CSAD for \textit{up} and CSAD for \textit{down} markets in each community. This will be explored in future works. Instead, we focus solely on determining whether the communities reveal herding behavior or not. To this end, we consider a community to exhibit herding when either CSAD or CSAD for \textit{up} market condition or CSAD for \textit{down} market condition reveals herding.

\begin{figure}[H]
\centering
\includegraphics[width = \textwidth]{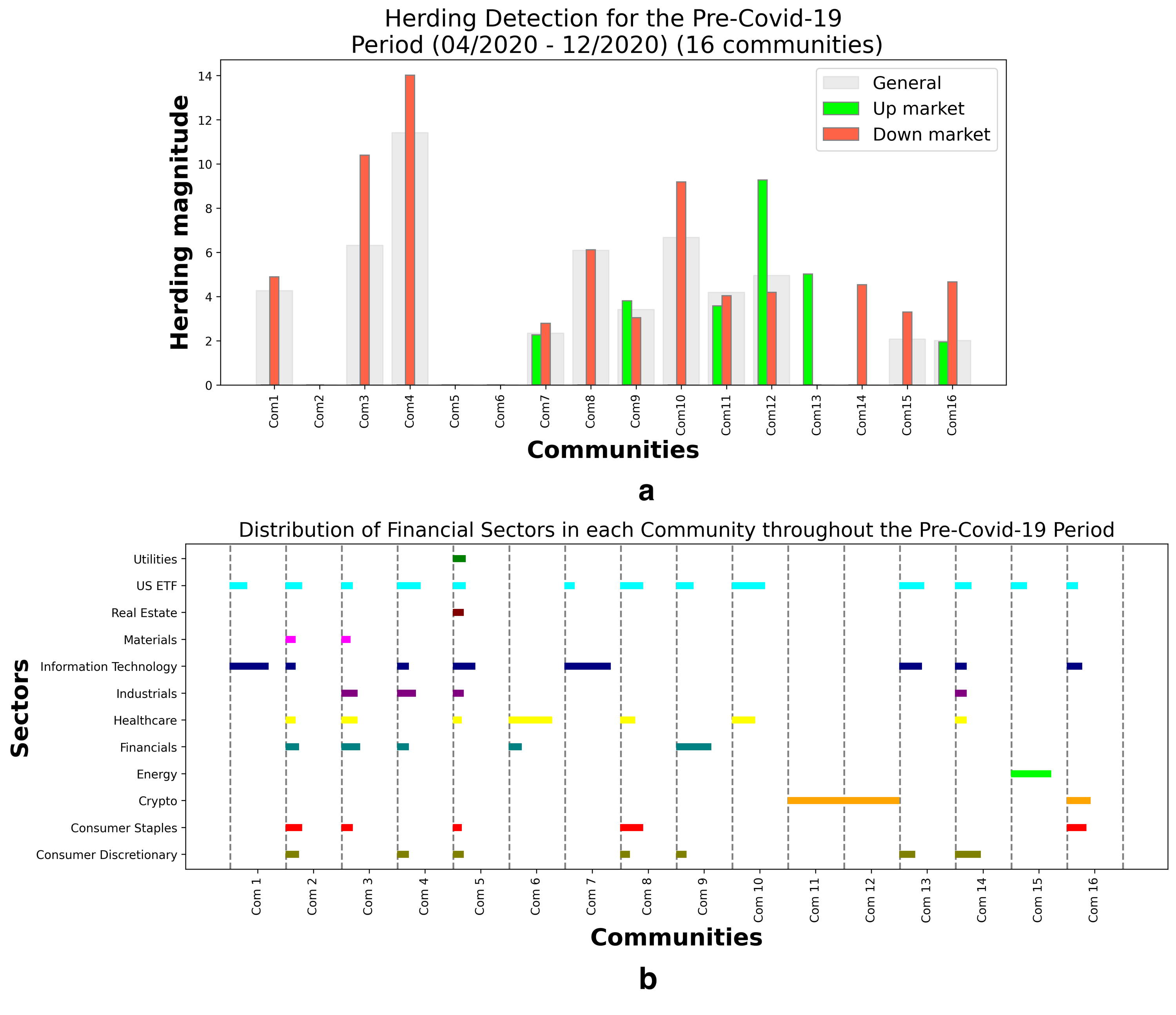}
\caption{\bf Herding detection results (a) and sector distribution (b) of each community during the Pre-Covid-19 period. More information on each sub-figure can be seen at \nameref{SI_Text2}. \label{Fig9}}
\end{figure}

\begin{figure}[H]
\centering
\includegraphics[width = \textwidth]{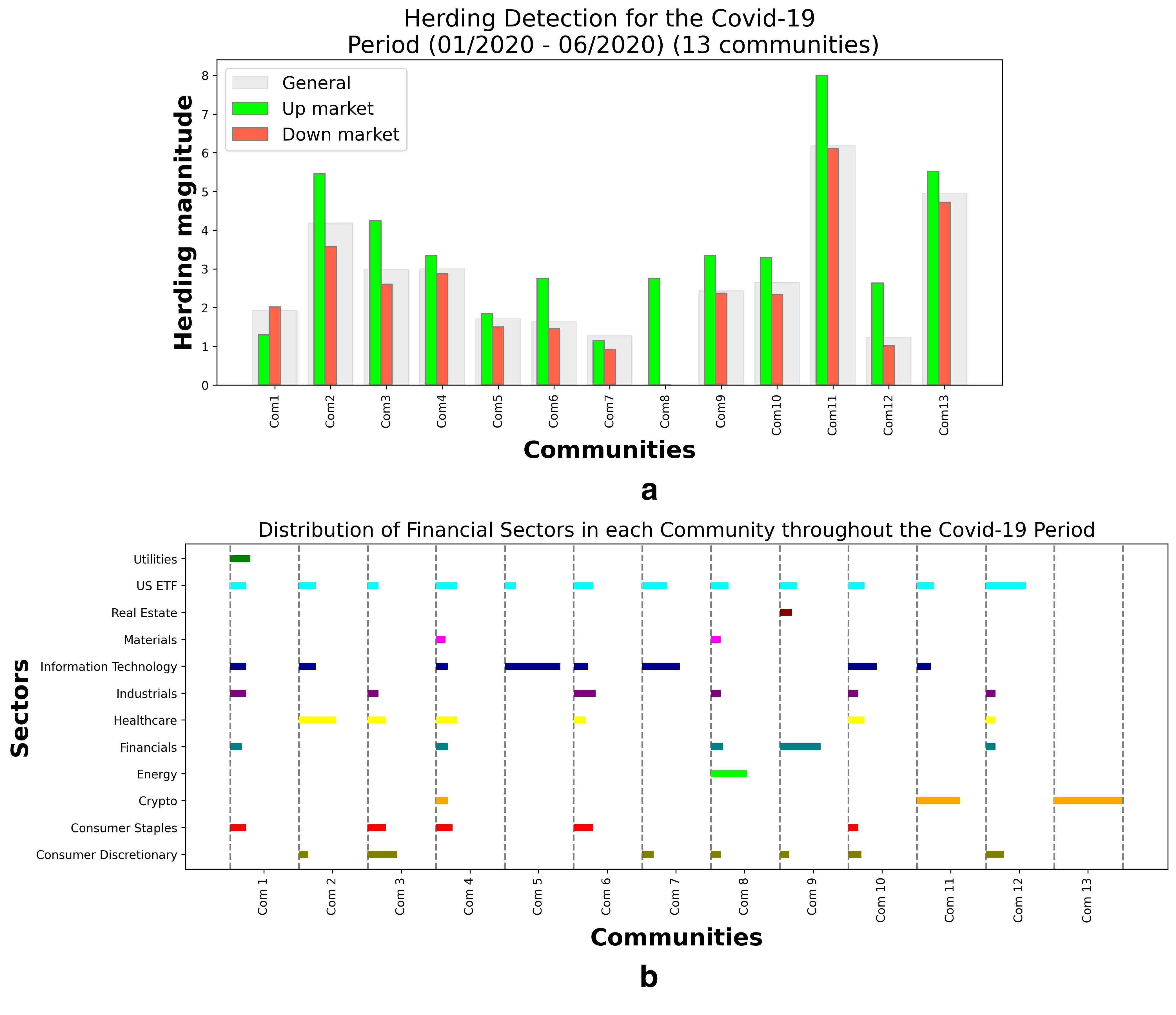}
\caption{ \bf Herding detection results (a) and sector distribution (b) of each community during the Covid-19 Pandemic period. More information on each sub-figure can be seen at \nameref{SI_Text2}. \label{Fig10}}
\end{figure}

\begin{figure}[H]
\centering
\includegraphics[width = \textwidth]{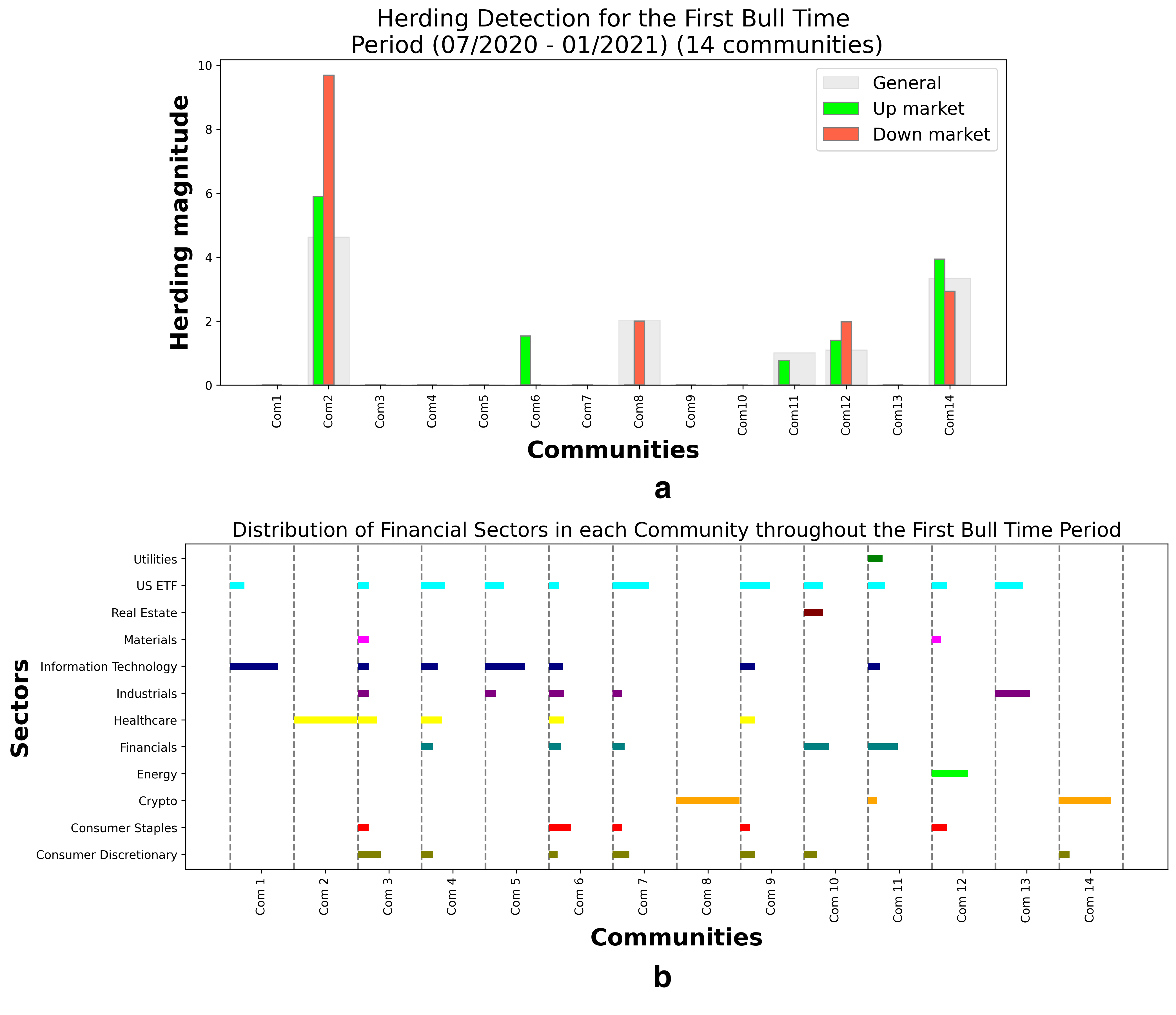}
\caption{\bf Herding detection results (a) and sector distribution (b) of each community during the first Bull Time period. More information on each sub-figure can be seen at \nameref{SI_Text2}. \label{Fig11}}
\end{figure}

\begin{figure}[H]
\centering
\includegraphics[width = \textwidth]{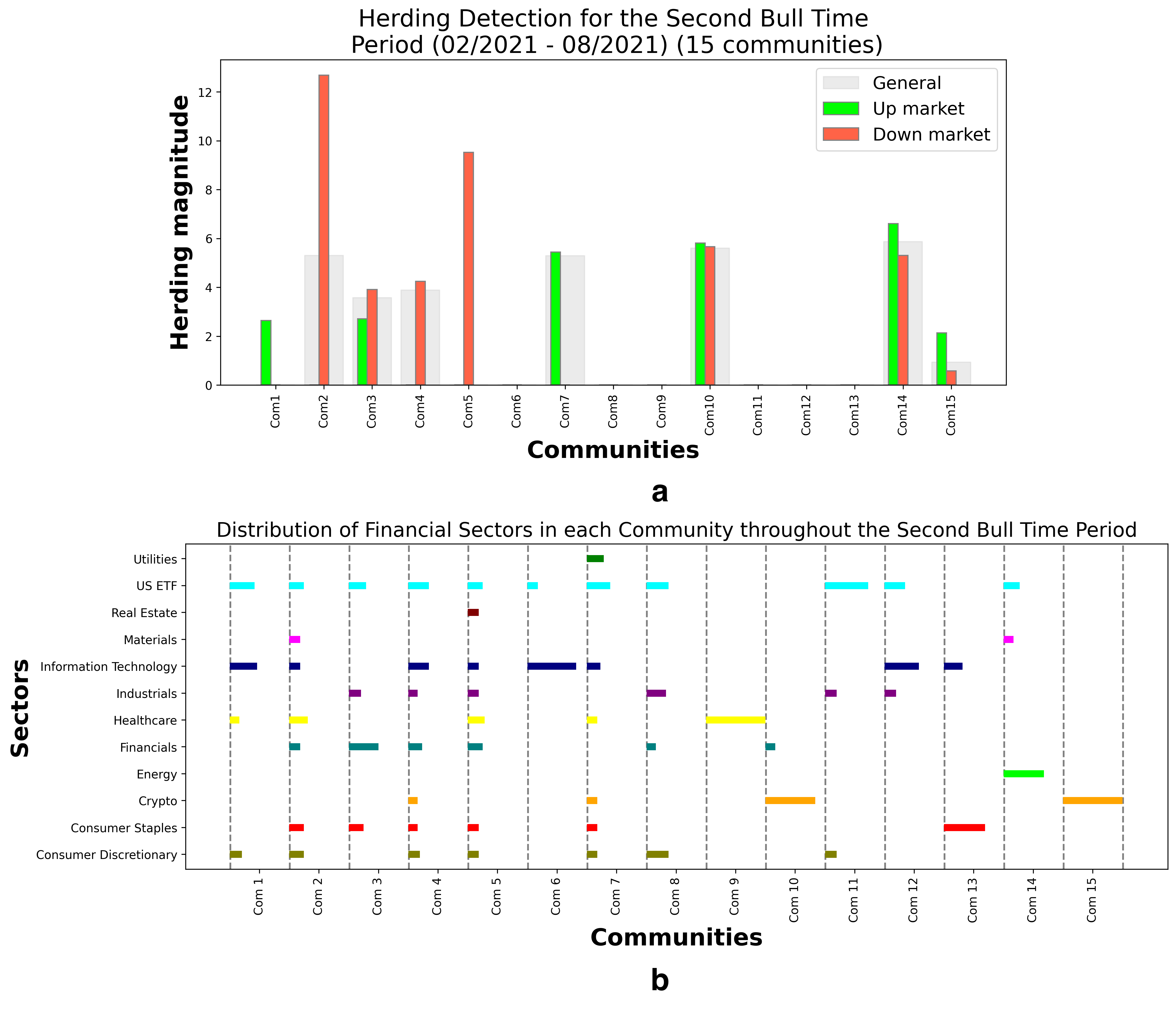}
\caption{\bf Herding detection results (a) and sector distribution (b) of each community during the second Bull Time period. More information on each sub-figure can be seen at \nameref{SI_Text2}. \label{Fig12}}
\end{figure}

\begin{figure}[H]
\centering
\includegraphics[width = \textwidth]{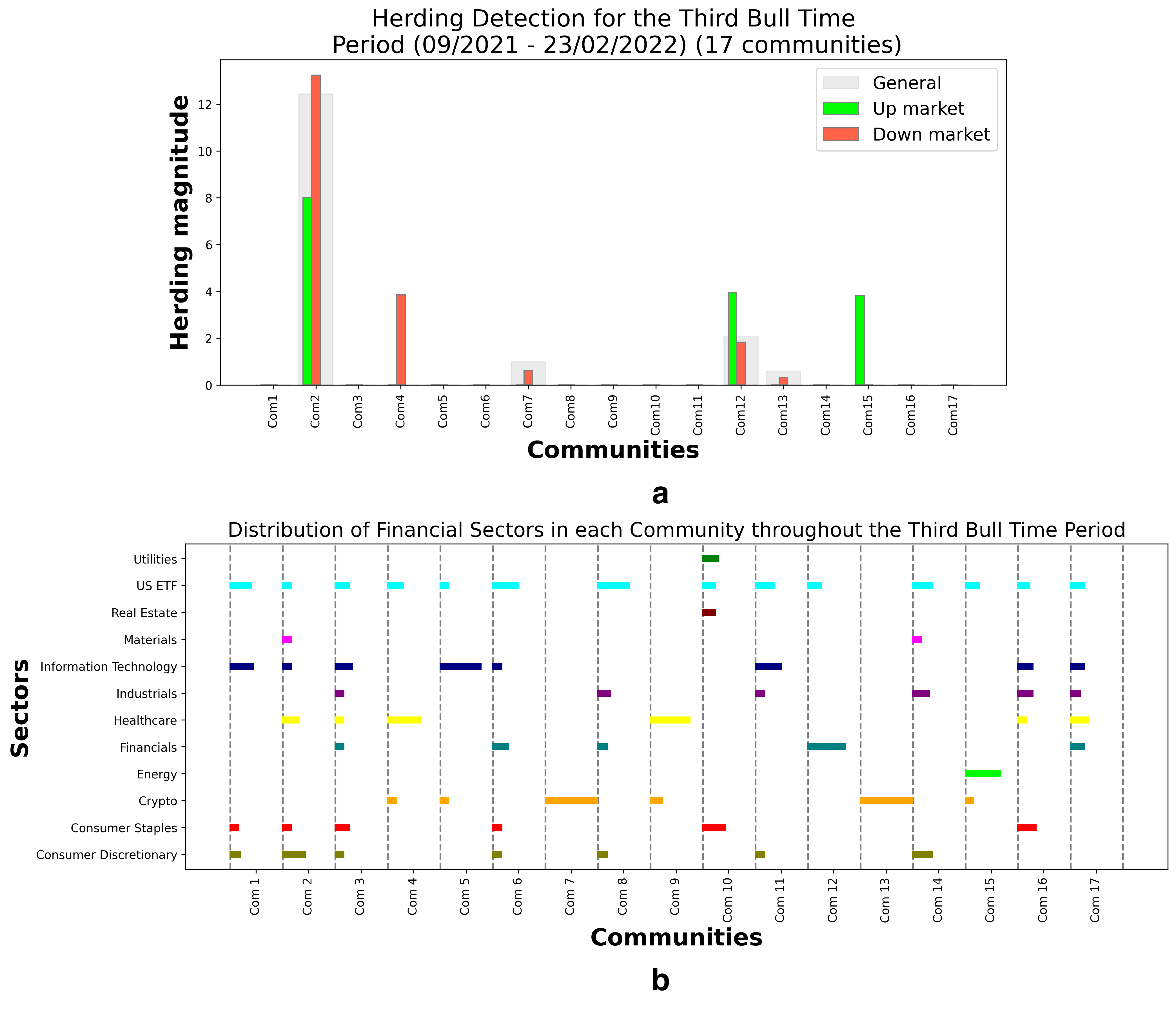}
\caption{\bf Herding detection results (a) and sector distribution (b) of each community during the third Bull Time period. More information on each sub-figure can be seen at \nameref{SI_Text2}. \label{Fig13}}
\end{figure}

\begin{figure}[H]
\centering
\includegraphics[width = \textwidth]{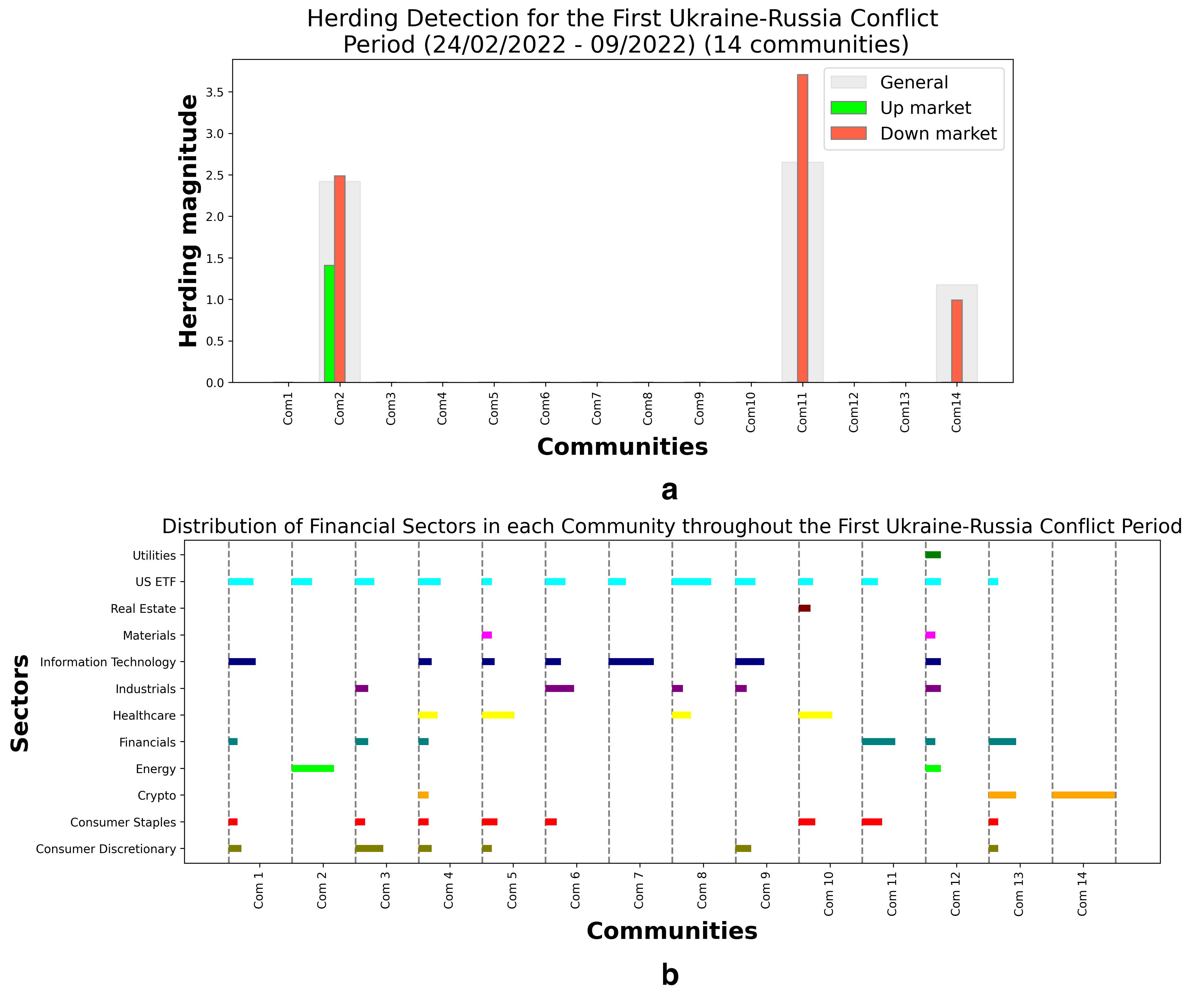}
\caption{\bf Herding detection results (a) and sector distribution (b) of each community during the first Ukraine-Russia Conflict period. More information on each sub-figure can be seen at \nameref{SI_Text2}. \label{Fig14}}
\end{figure}

\begin{figure}[H]
\centering
\includegraphics[width = \textwidth]{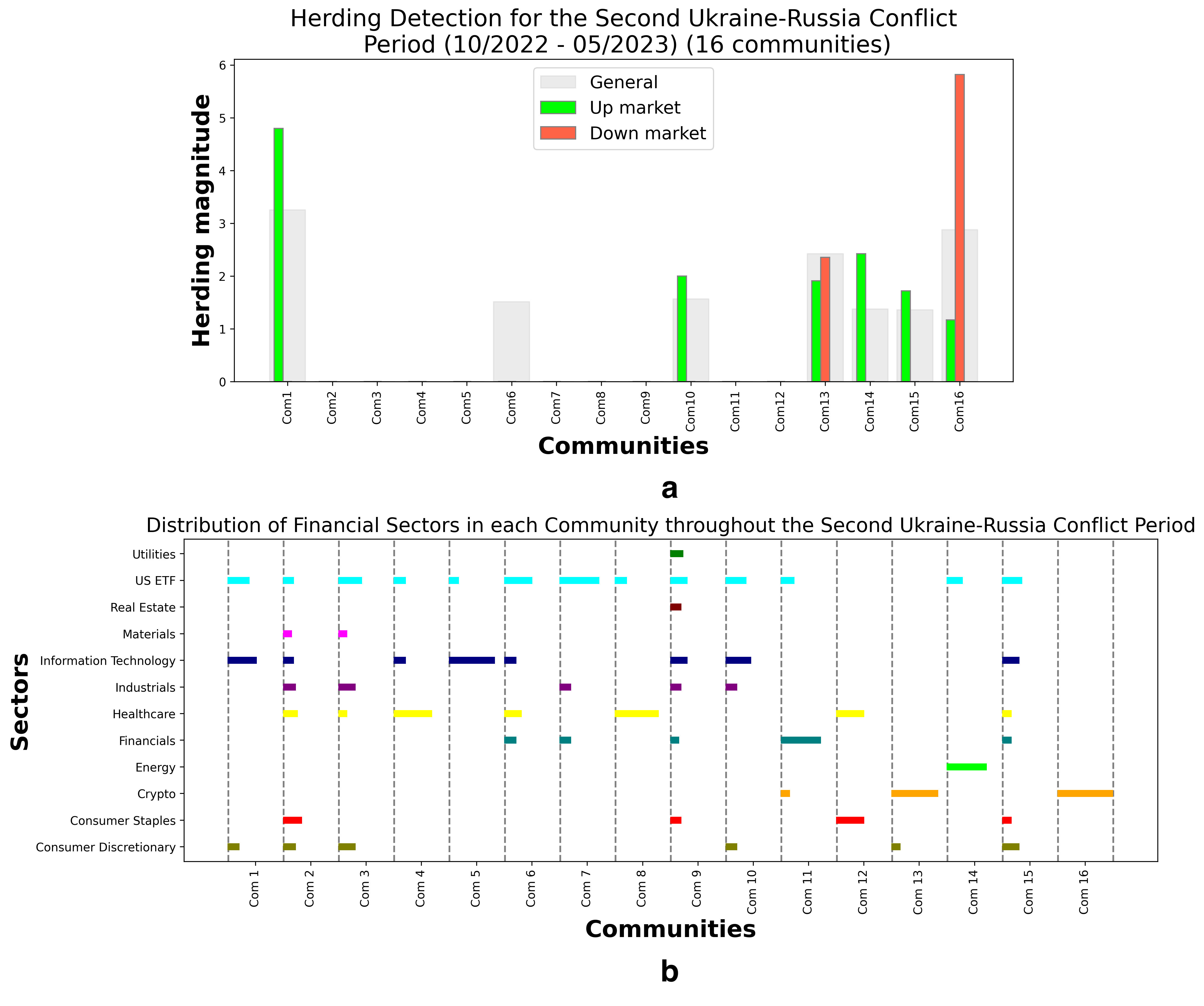}
\caption{\bf Herding detection results (a) and sector distribution (b) of each community during the second Ukraine-Russia Conflict period. More information on each sub-figure can be seen at \nameref{SI_Text2}. \label{Fig15}}
\end{figure}

Figs \ref{Fig9}-\ref{Fig15} illustrate the herding behavior results (Figs \ref{Fig9}a-\ref{Fig15}a) and sector distribution (Figs \ref{Fig9}b-\ref{Fig15}b) of each community in each sub-period.  The assets in each community are listed in \nameref{S4_Tab}.  At first glance, it is clear that the first two sub-periods look rather similar in terms of the density of herding signals, as most communities reveal herding behavior in Figs \ref{Fig9}a and \ref{Fig10}a. This is contrary to the last five sub-periods (Figs \ref{Fig11}a-\ref{Fig15}a) since these predominantly display a dearth of herding. Perhaps, the high density of herding during the pre-and the peak of the pandemic was caused by two major events, as mentioned previously, the US-China trade war throughout 2019 and the contagious disease Covid-19 which started early in 2020, respectively. Whereas, once the pandemic started to be controlled and the global economy recovered from the loss, financial markets resumed their normal behavior. Thus, investors stopped mimicking each other and began to make their own objective decisions, and the herding was lessened accordingly.

We notice from Figs \ref{Fig9}b-\ref{Fig15}b that cryptocurrencies are close to each other and tend to form 2 to 3 large communities among themselves. Observing these communities throughout different sub-periods, there is evidence of herding most of the time, as shown in Figs \ref{Fig9}a-\ref{Fig15}a. Compared to the herding detection results in the cryptocurrency market displayed in Tables \ref{tab:4.1} and \ref{tab:4.2}, it is clear that without critical events impacting the entire cryptocurrency market, herding still exists in most cryptocurrencies, but at a community level, i.e herding behaviour is pronounced in different subsets of cryptocurrencies, while there was an absence of herding signals when looking at all cryptocurrencies at once. It is challenging to find the reasons causing the herding in such cases since there was no major event that could influence the market~\cite{man22,rai22,jeo24}. However, this market seems to be mainly driven by investors' emotions~\cite{nae21} and accessible news on online platforms~\cite{wak22}. In addition, cryptocurrency investors tend to adopt a short-term investment strategy and depend on price movements of the coins rather than reliable factors that are seen in traditional markets~\cite{jal20}. Consequently, the herding in those cryptocurrency communities could arise from either the potential profits or the fear of losing money at that time, encouraging investors to buy or sell. Moreover, investors might be aware of the co-movements between different cryptocurrencies. Thus, when the opportunity to gain profits (or the selling time) arrives, they tend to buy (or sell) the coins that have the same price movements to get higher profits (or mitigate the loss). This could explain why cryptocurrencies mainly formed 2 separate communities in the Pre-Covid-19 (Fig \ref{Fig9}) and the third Bull Time (Fig \ref{Fig13}) sub-periods, with these communities revealing herding behavior, while there was no herding when we detected the whole cryptocurrency market in the previous section.

Regarding other communities, each of them has a mixture of US ETFs and stocks where the US ETFs tend to act as a basket of most stocks in that community. The herding in these communities follows the same rule as the cryptocurrency communities since each sub-period reveals significant herding behavior in several communities. Remarkably, we obtained several valuable findings from the herding results of these stocks and US ETFs. In particular, there are 4 stock sectors (Energy, Healthcare, Information Technology and Financials) in which for each sector, its assets and relevant ETFs tend to form communities among themselves, throughout the examined period. Moreover, the herding signals found in each of these sectors coincide with the timing of economic and political events that specifically influence that sector.
\begin{itemize}
    \item \textbf{Energy Sector}: In our case, the energy stocks we study mainly consist of petroleum treatment companies and petroleum exploiting companies. Thus, the herding found in this study is primarily driven by oil-related events. We found that all energy stocks and ETFs are exposed to herding across all sub-periods. Energy companies, in general, have consistently experienced tough times over the past 4 years from 2019 until the present. During this time, they were hit by a series of different events, causing large fluctuations in their corresponding stock prices over time, including the US-China trade war, the Covid-19 pandemic, the petroleum oversupply due to the shale revolution~\cite{nat20,bp20}, the Russia-Saudi Arabia price war~\cite{ma21,fat20} and the Ukraine-Russia war~\cite{tee23}.
    
    \item \textbf{Technology Sector}: Technology stocks are grouped into several communities and reveal herding behavior in four sub-periods, including Pre-Covid-19 (coinciding with the US-China trade war~\cite{kwa20,bow20,hou20}), Covid-19 Pandemic (coinciding with the downturn time caused by the pandemic followed by a global economic recession), the second Bull Time (coinciding with the price surge in financial markets\cite{wor21})  and the second Ukraine-Russia Conflict (coinciding with the AI boom period that gave an impetus to the Technology sector~\cite{wea23,gui23}).

    \item \textbf{Healthcare Sector}: Herding behavior was found in most healthcare stocks and ETFs during three sub-periods: the Covid-19 Pandemic,the first and third Bull Time. This herding behavior likely stems from the great attraction of investors toward pharmaceutical companies due to their positive performance during the pandemic. In particular, dozens of pharmaceutical companies began vaccine research from early 2020 due to the high demand for vaccinations for the entire world~\cite{and20,gas22,bow22}. Moreover, other pharmaceutical companies, though not participating in producing vaccines, produced critical elements that were necessary for vaccine production such as nanoparticles or medical devices for hospitals and daily usage such as the Covid-19 test kits. These contributions also pushed their performance during this time~\cite{bow22,ami23}.

    \item \textbf{Financial Sector}: In this study, the financial sector mainly consists of banking firms. We found that this industry was exposed to herding behavior in the first six sub-periods, except for the second Ukraine-Russia Conflict time. The financial sector is strongly related to economic conditions and reflects the performance of the economy, especially the banking firms since they have the responsibility of preserving the economic stability of a country and also keeping the country away from financial crises through various schemes like reducing interest rates~\cite{amp22}, imposing new monetary policy comes from FED~\cite{bat23} and cutting companies' liquidity risks~\cite{gor20}. Consequently, economic uncertainty forces the financial sector to experience distress. Indeed, the global economy had been experiencing ups and downs continually through out the herding periods in this stock sector.

\end{itemize}

For the remaining sectors (i.e. Industry, Consumer Discretionary and Consumer Staples), we found that assets within the same sector never group into a large community in all sub-periods, even during the Covid-19 pandemic. Instead, they are broken down and distributed almost equally into several communities. This result indicates that investors did not perform herding on a sector basis in these sectors, even during major events. Furthermore, we acknowledge that some communities are mixed by multiple sectors and reveal herding, but it is unclear why the herding behavior exists in those cases as there seems not to be a rule/convention for these cases. However, most of such communities appear during the economic uncertainty times, i.e. during the trade war (Fig \ref{Fig9}) and the Covid-19 pandemic (Fig \ref{Fig10}). By contrast, this phenomenon was not pronounced from the first Bull Time sub-period onwards (Figs \ref{Fig11}- \ref{Fig15}). Thus, it seems to be more likely that negative market conditions resulted in this phenomenon. 
However, since we don't have a confident justification for this phenomenon, we will leave this result for future work.

From this experimental result, we can conclude that herding behavior should be observed from two perspectives, one from the entire market and the other from separate subsets (communities) within the market. This is because different perspectives provide us with different aspects of the market. Specifically, herding behavior occurs widely in most assets of a particular type of investment vehicle when there is a critical event that directly impacts the market. Hence, most assets involved are influenced and thus investors who are trading those assets tend to perform herding as a reaction to the event. This is why we found the herding in traditional markets during the US-China trade war, the herding in all markets during the peak of the Covid-19 pandemic and the herding in the cryptocurrency market during bull market conditions, while the remaining sub-periods tend to show no evidence of herding (Tables \ref{tab:4.1} and \ref{tab:4.2}). By contrast, we found that herding behavior occurs constantly in all types of investment vehicle but at a smaller scale, i.e. community level, as shown in Figs \ref{Fig9}-\ref{Fig15}. In addition, herding behavior found in these communities mainly comes from specific events that solely impact that community. Subsequently, although herding might not exist in a whole market, it can be found by looking at a particular community/sector.

\subsection{Contagion Effects between Cryptocurrencies, Stocks and US ETFs}
Following the above herding detection results, we conduct another experiment examining the effects of the phenomenon of financial contagion between the assets. This experiment aims to identify the shock drivers between cryptocurrencies, stocks and US ETFs. Since herding has been identified as a cause of the contagion~\cite{cip08}, understanding the patterns of contagion over time helps identify the herding transmitters among these assets. Additionally, other authors have also suggested as a link between ETFs and underlying assets, the mechanism of \textit{long the stock, short the ETF} arbitrage by Hedge funds~\cite{shi21}, which tends to amplify the contagion effects between the assets and to give an impetus to herding behavior among investors on not only the ETFs but also their underlying assets.

Based on the 30-min dataset used in this study, we estimate the 30-min lag contagion effects between 222 assets in each of the seven sub-periods using the VAR model. The results are saved in tables and displayed in \nameref{S1_File} where each column represents a shock receiver and each row represents a shock transmitter, a value that is not 0.0 indicates evidence of contagion from one asset to another asset.

To facilitate the analysis of these tables, we visualize them as 3D figures. In particular, we transform each contagion table result (corresponding to one sub-period) into six figures, including:
\begin{itemize}
    \item[1)] The positive contagion from each of 222 assets to cryptocurrencies.
    \item[2)] The negative contagion from each of 222 assets to cryptocurrencies.
    \item[3)] The positive contagion from each of 222 assets to stocks.
    \item[4)] The negative contagion from each of 222 assets to stocks.
    \item[5)] The positive contagion from each of 222 assets to US ETFs.
    \item[6)] The negative contagion from each of 222 assets to US ETFs.
\end{itemize}

\begin{figure}[H]
\centering
\includegraphics[width = \textwidth]{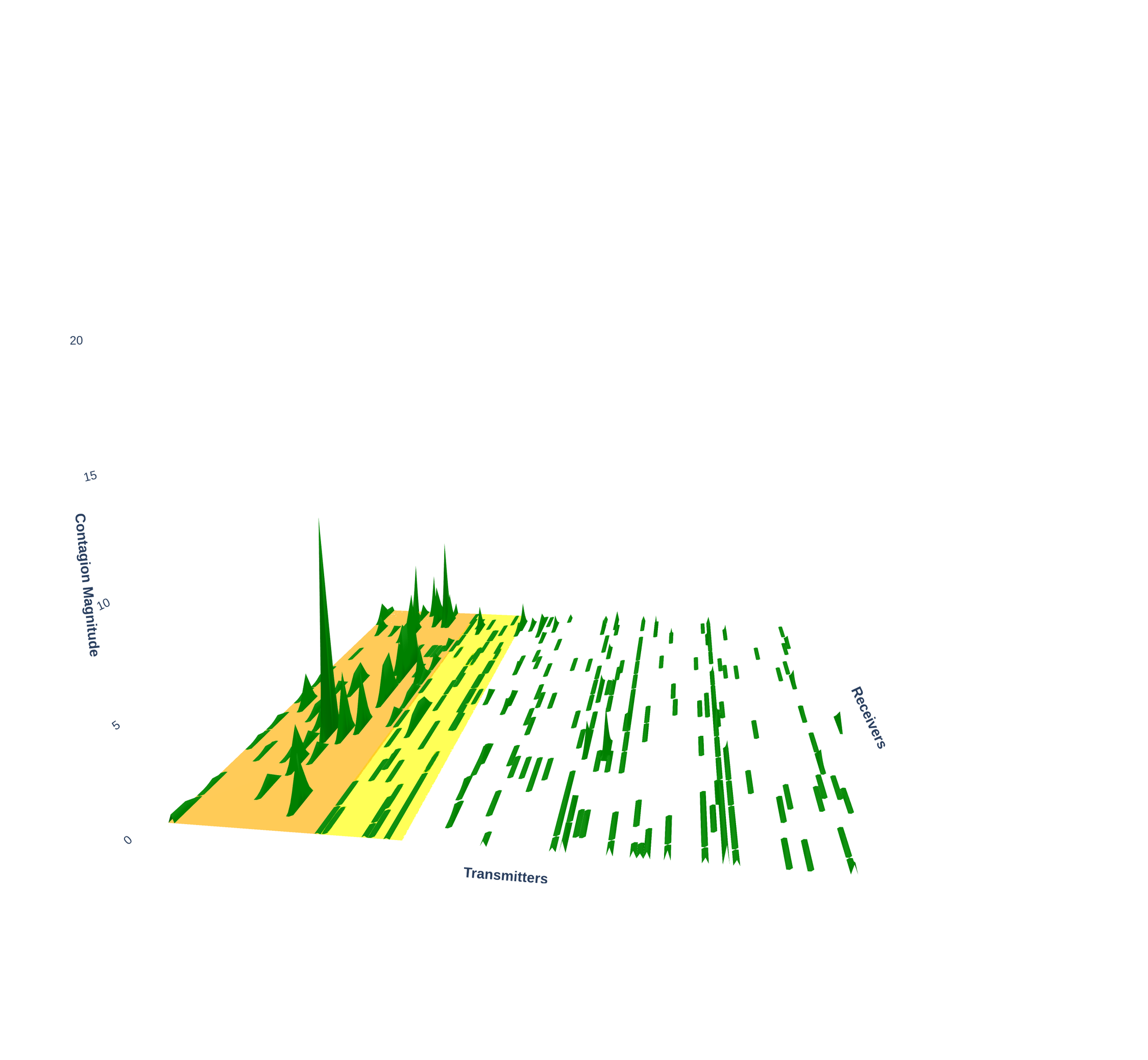}
\caption{\bf The positive contagion effects from each of 222 assets (y-axis) to cryptocurrencies (x-axis) during the Pre-Covid-19 sub-period. The yellow area represents cryptocurrencies (as shock transmitters) and the orange area represents US ETFs (as shock transmitters). The green spikes represent evidence of positive contagion from one asset to one cryptocurrency. \label{Fig16}}
\end{figure}

\begin{figure}[H]
\centering
\includegraphics[width = \textwidth]{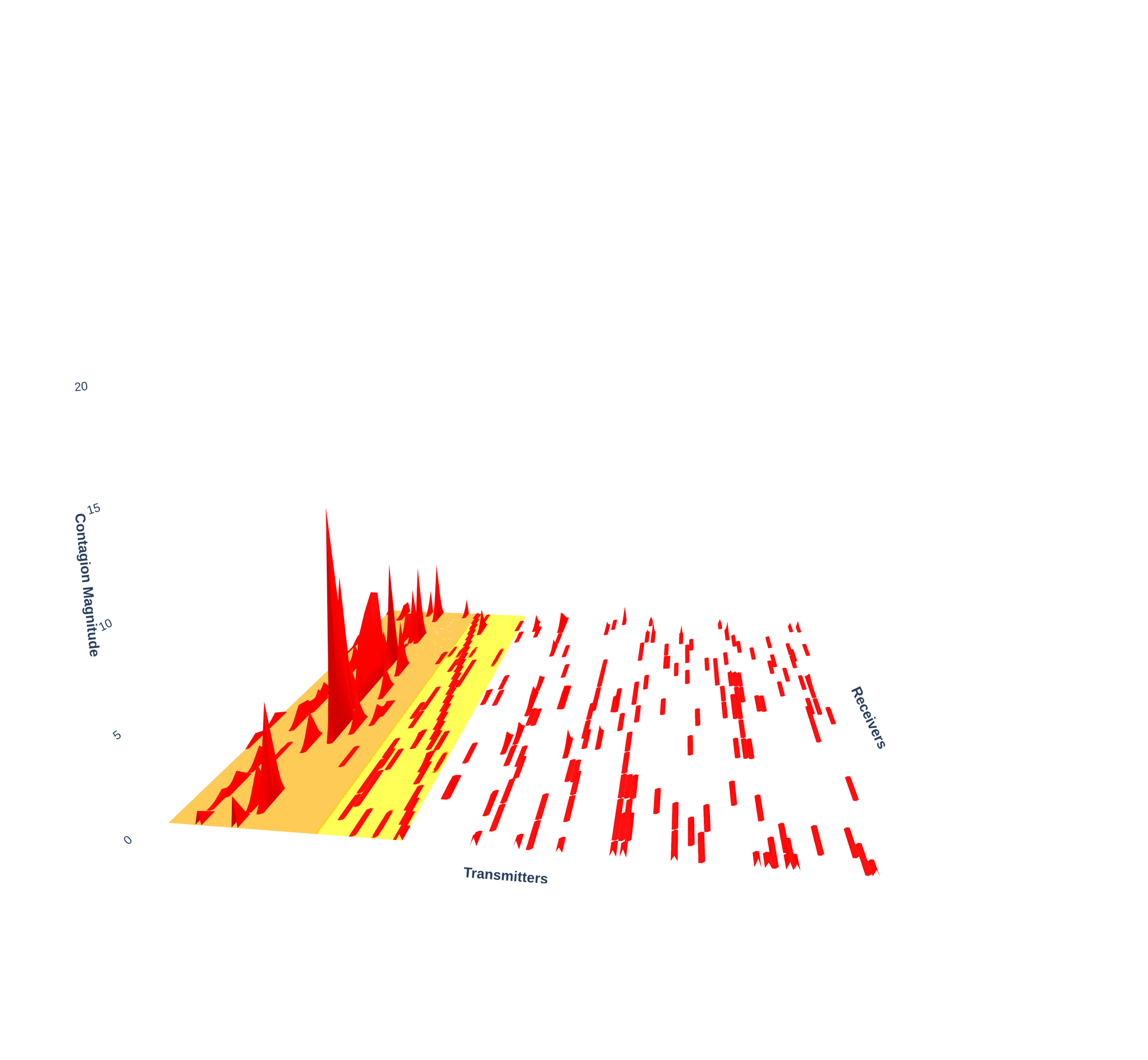}
\caption{\bf The negative contagion effects (absolute values) from each of 222 assets (y-axis) to cryptocurrencies (x-axis) during the Pre-Covid-19 sub-period. The yellow area represents cryptocurrencies (as shock transmitters) and the orange area represents US ETFs (as shock transmitters). The red spikes represent evidence of negative contagion from one asset to one cryptocurrency. \label{Fig17}}
\end{figure}

\begin{figure}[H]
\centering
\includegraphics[width = \textwidth]{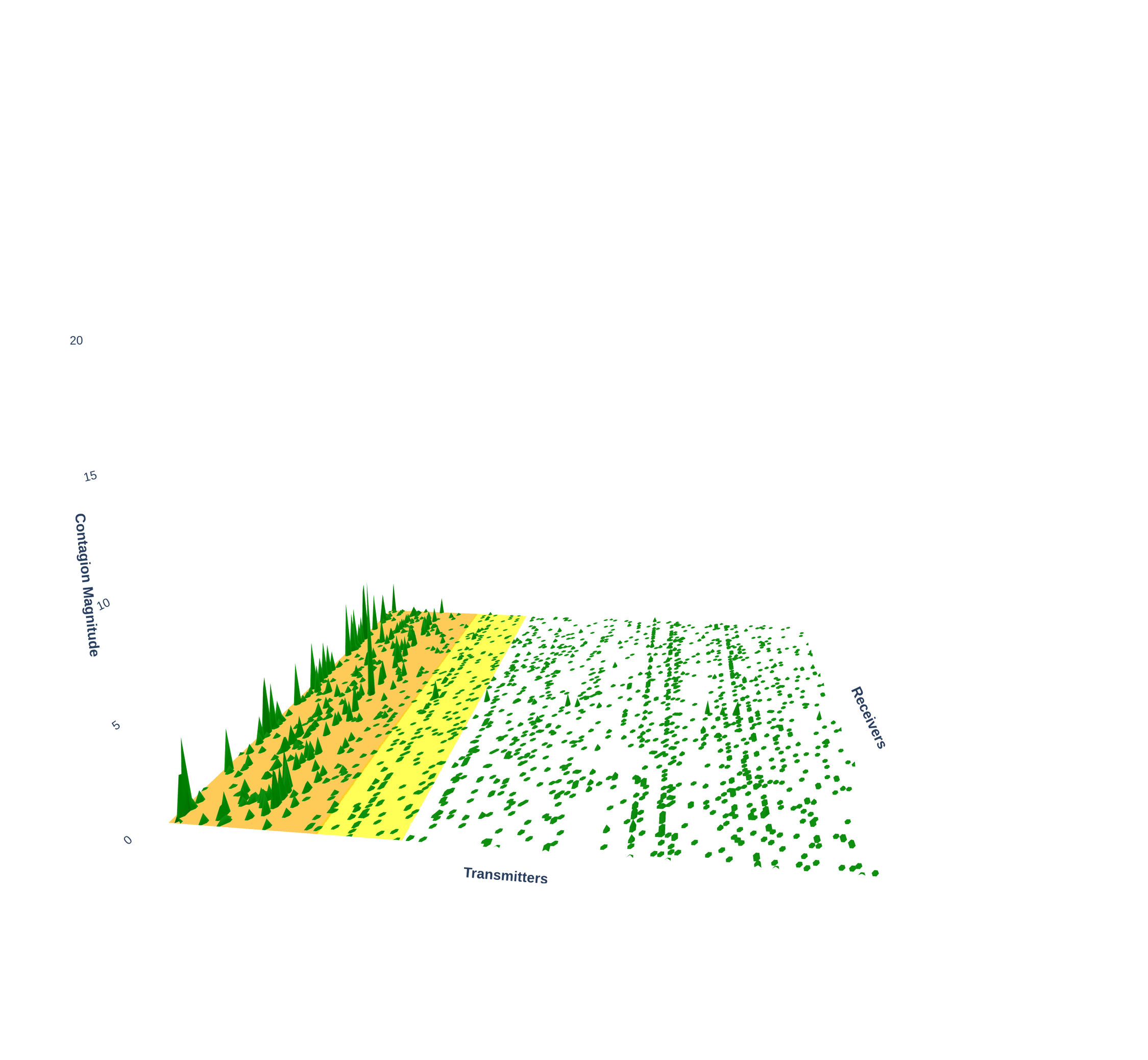}
\caption{\bf The positive contagion effects from each of 222 assets (y-axis) to stocks (x-axis) during the Pre-Covid-19 sub-period. The yellow area represents cryptocurrencies (as shock transmitters) and the orange area represents US ETFs (as shock transmitters). The green spikes represent evidence of positive contagion from one asset to one stock. \label{Fig18}}
\end{figure}

\begin{figure}[H]
\centering
\includegraphics[width = \textwidth]{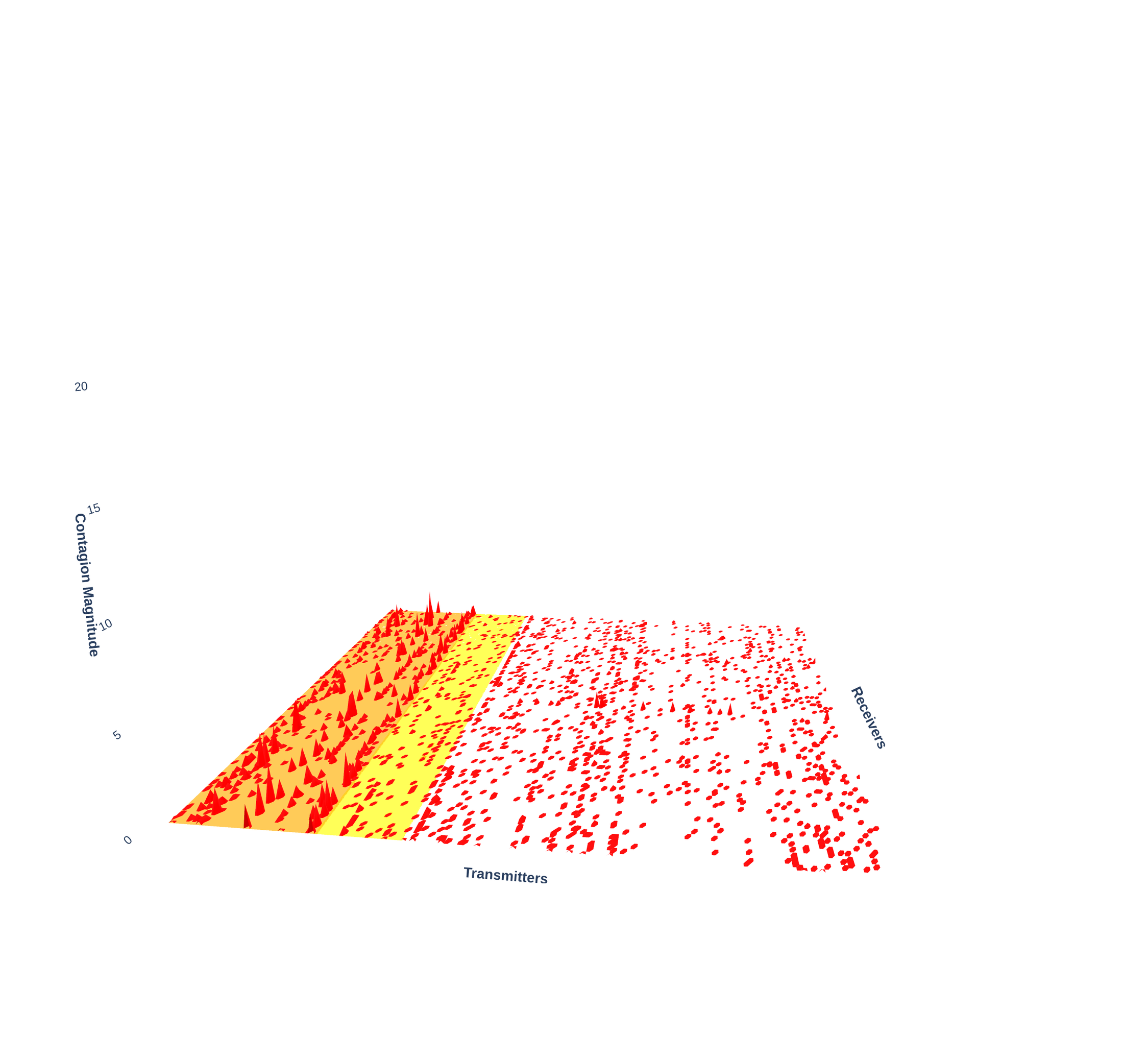}
\caption{\bf The negative contagion effects (absolute values) from each of 222 assets (y-axis) to stocks (x-axis) during the Pre-Covid-19 sub-period. The yellow area represents cryptocurrencies (as shock transmitters) and the orange area represents US ETFs (as shock transmitters). The red spikes represent evidence of negative contagion from one asset to one stock. \label{Fig19}}
\end{figure}

\begin{figure}[H]
\centering
\includegraphics[width = \textwidth]{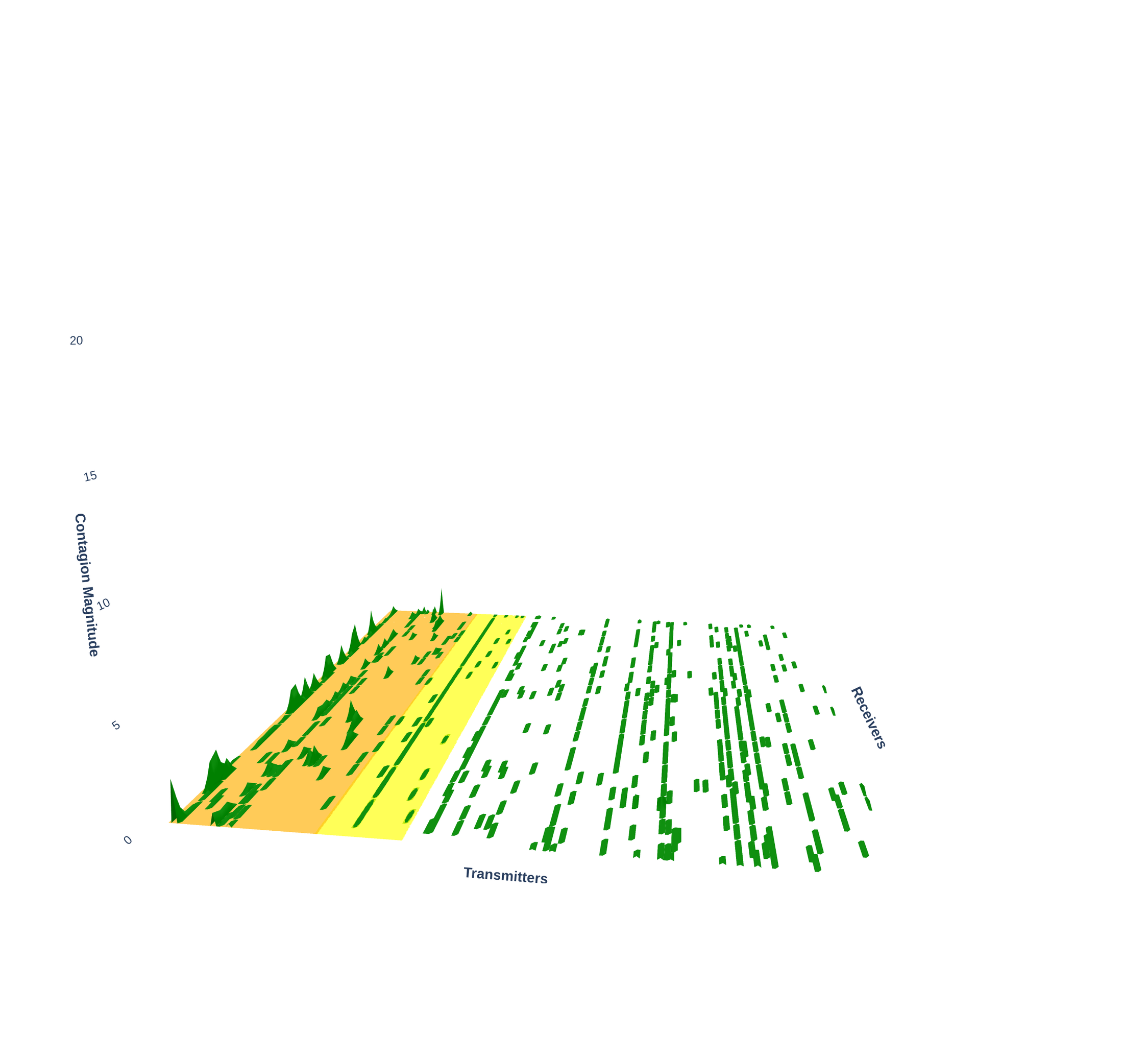}
\caption{\bf The positive contagion effects from each of 222 assets (y-axis) to US ETFs (x-axis) during the Pre-Covid-19 sub-period. The yellow area represents cryptocurrencies (as shock transmitters) and the orange area represents US ETFs (as shock transmitters). The green spikes represent evidence of positive contagion from one asset to one US ETF. \label{Fig20}}
\end{figure}

\begin{figure}[H]
\centering
\includegraphics[width = \textwidth]{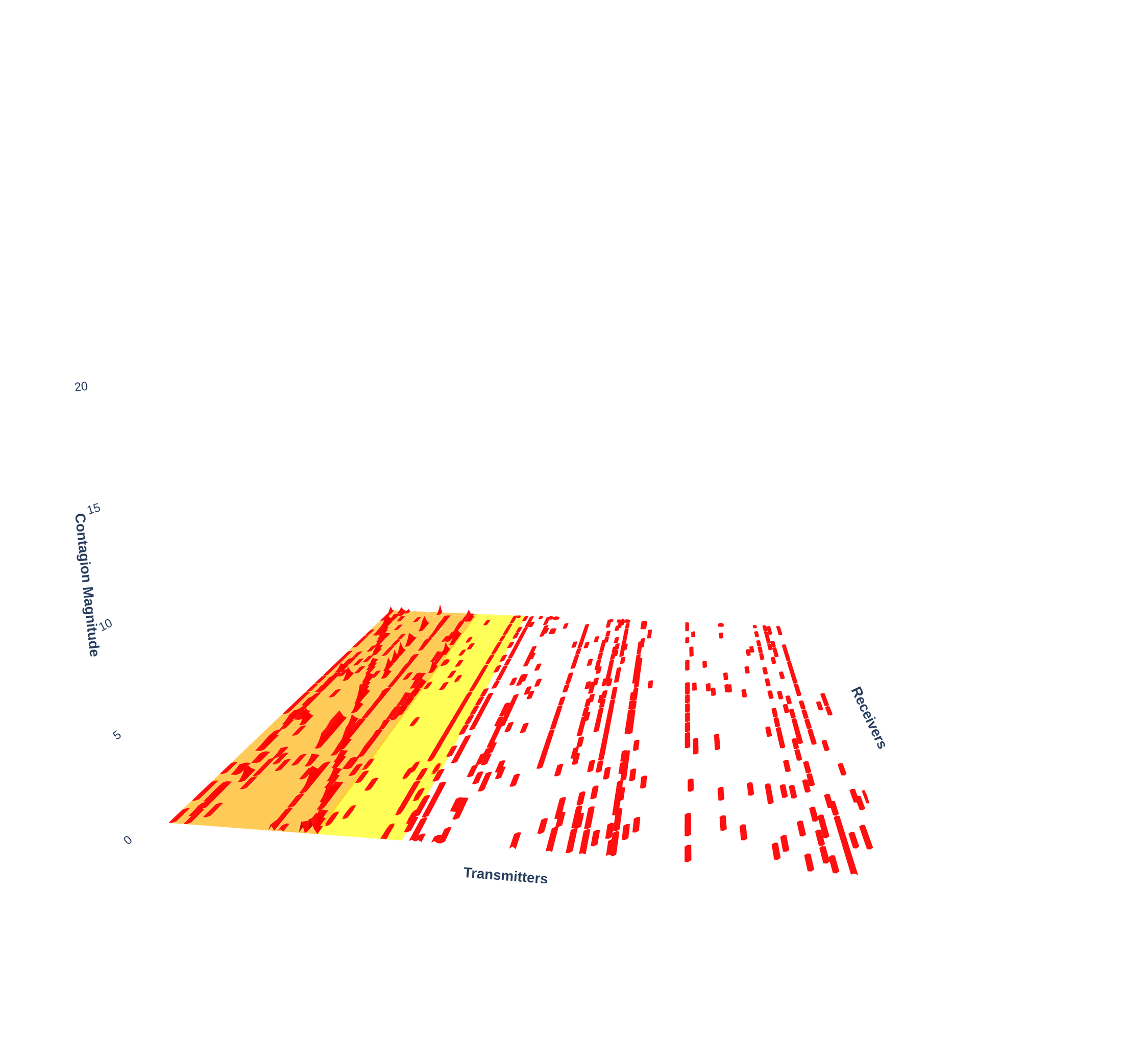}
\caption{\bf The negative contagion effects (absolute values) from each of 222 assets (y-axis) to US ETFs (x-axis) during the Pre-Covid-19 sub-period. The yellow area represents cryptocurrencies (as shock transmitters) and the orange area represents US ETFs (as shock transmitters). The red spikes represent evidence of negative contagion from one asset to one US ETF. \label{Fig21}}
\end{figure}

Examining multiple assets from each type of investment vehicle, rather than focusing only on selected representatives provides a broader view of the contagion effects between different types of investment vehicle. Specifically, the contagion exists widely within each pair of categories, i.e. stocks and US ETFs, cryptocurrencies and US ETFs, and stocks and cryptocurrencies, as several assets in one category transmit shocks to those in another. However, we found that US ETFs are always the strongest shock transmitters to other types of investment vehicle and also among themselves, across different sub-periods. This discloses the strong influence of US ETFs not only on stocks buton  also cryptocurrencies, regardless of the existence of economic crises or not. By contrast, although many cryptocurrencies act as shock transmitters to stocks and US ETFs, these contagion effects are generally weak. Without loss of generalization, we display 3D figures of contagion effects between the assets during the Pre-Covid-19 sub-period in Figs \ref{Fig16}-\ref{Fig21}. For each figure, the x-axis represents the assets acting as shock receivers, the y-axis represents the assets acting as shock transmitters, and the z-axis represents the contagion magnitude (absolute value of the contagion coefficient) from one asset to another.

To explore further the contagion results, we calculate the average contagion magnitude between the three types of investment vehicle considered. In particular, we measure the average contagion magnitude that a type of investment vehicle transmits to other types of investment vehicle and also itself. For example, the value of this metric from stocks to cryptocurrencies is calculated in 2 steps. Firstly, averaging all the absolute contagion coefficients from stocks to each of the cryptocurrencies. After this step, each cryptocurrency has an average contagion magnitude that the stock market transmits to it. Secondly, we average the average values that the stock market transmits to each of the cryptocurrencies.

As shown in Table \ref{tab:5}, the contagion from US ETFs to cryptocurrencies is consistently strong and higher than that for other transmissions. Moreover, another strong contagion effect is also seen from US ETFs to stocks. These results remain unchanged across seven sub-periods, indicating a strong influence of US ETFs on both cryptocurrencies and stocks, with a particularly pronounced effect on cryptocurrencies. Similarly, cryptocurrencies are also greatly influenced by the stock market, as indicated by a relatively high contagion magnitude from stocks to cryptocurrencies, although this effect is weaker compared to the influence of US ETFs. In contrast, contagion from cryptocurrencies to traditional assets (both stocks and US ETFs), while present, is negligible across all sub-periods, highlighting the limited influence of cryptocurrencies on these markets. In the case of stocks, they also exhibit only modest contagion effects on US ETFs. Instead, US ETFs primarily transmit shocks among themselves, indicating their strong internal contagion. Notably, contagion effects between these types of investment vehicles peak during the Covid-19 pandemic sub-period, as shown by a rise in most contagion coefficients during this time. After this period, however, these contagion levels declined.
\begin{table}[h!]

\begin{adjustwidth}{-0.4in}{0in}
\caption{\bf Average Contagion Magnitude from one type of investment vehicle to the others and itself. Three types of investment vehicle are considered, namely cryptocurrencies, stocks and US ETFs}
\begin{tabular}{cccccccc}
\hline
Transmission & Pre-Covid-19 & Covid-19 & Bull Time 1 & Bull Time 2 & Bull Time 3 & U-R Conflict 1 & U-R Conflict 2 \\ \hline
Crypto to crypto & 0.1674 & 0.3536 & 0.2391 & 0.3040 & 0.4308 & 0.1592 & 0.5855 \\
Stock to stock   & 0.0962 & 0.1779 & 0.1250 & 0.1201 & 0.1491 & 0.1438 & 0.1377 \\
US ETF to US ETF & 0.3191 & 0.9974 & 0.4323 & 0.4951 & 0.4279 & 0.4336 & 0.4974 \\
Stock to crypto  & 0.4012 & 0.3823 & 0.4924 & 0.8898 & 0.4939 & 0.4088 & 0.3469 \\
US ETF to crypto & 1.7424 & 2.2098 & 2.3056 & 3.1864 & 2.8145 & 4.7943 & 1.3000 \\
Crypto to stock  & 0.0459 & 0.1225 & 0.0783 & 0.0434 & 0.1264 & 0.1244 & 0.0695 \\
US ETF to stock  & 0.5077 & 1.0208 & 0.6153 & 0.5964 & 0.7415 & 0.4669 & 0.5632 \\
Crypto to US ETF & 0.0215 & 0.0873 & 0.0293 & 0.0259 & 0.1004 & 0.0811 & 0.0556 \\
Stock to US ETF  & 0.0628 & 0.1562 & 0.0910 & 0.1184 & 0.1091 & 0.1213 & 0.1098 \\ \hline
\end{tabular} \label{tab:5}
\end{adjustwidth}
\end{table}

Linking to the herding behavior discussed in previous sections, US ETFs appear to be one of the herding drivers in financial markets, particularly for cryptocurrencies. In other words, similar trading actions between investors in not only cryptocurrency but also traditional markets may be influenced by the movements of US ETFs, reinforcing the findings of \cite{cip08}. Additionally, US ETFs seem to exhibit self-reinforcing herding behavior, where they act as drivers of herding among themselves. This substantial amount of contagion (and so influence) from US ETFs to other assets reinforces their role as the central nodes of communities within a correlation network, as shown in the previous section. Moreover, when comparing contagion effects from stocks and cryptocurrencies to US ETFs, stocks generally exert a stronger influence on US ETFs than cryptocurrencies do. To this end, US ETFs and stocks exhibit a greater mutual influence compared to the influence between US ETFs and cryptocurrencies. This explains why US ETFs and stocks are more likely to belong to the same community, while cryptocurrencies tend to form communities between themselves, as discussed in the previous section.

Tables \ref{tab:6} and \ref{tab:7} shows the top 3 US ETFs in each sub-period that transmit their shocks to the highest number of cryptocurrencies and stocks, respectively. As can be seen, the most influential US ETFs tend to be different in different sub-periods, for both cryptocurrencies and stocks. However, SVXY (ProShares Short VIX Short-Term Futures ETF) is the most influential ETF on stocks on a regular basis, transmitting its shocks to the highest number of stocks in the Pre-Covid-19, the second Bull Time and the first Ukraine-Russia Conflict sub-periods. This characteristic is not true for the case of cryptocurrencies since the most influential US ETF varies over time. Furthermore, S\&P and Nasdaq-related ETFs tend to be common transmitters to both stocks and cryptocurrencies across different sub-periods. 

\begin{table}[h!]

\begin{adjustwidth}{-0.4in}{0in}
\caption{ \bf Top three US ETFs in each sub-period that transmit their shocks to the highest number of stocks}
\begin{tabular}{cccccccc}
\hline
Influential Order & Pre-Covid-19 & Covid-19 & Bull Time 1 & Bull Time 2 & Bull Time 3 & U-R Conflict 1 & U-R Conflict 2 \\ \hline
1 & SVXY & IYF & IDU  & SVXY & QTEC & SVXY & IWM  \\
2 & SLYG & QQQ & ONEQ & FDLO & IBB  & IJT  & FTXO \\
3 & IJK  & OEF & ITOT & QQQ  & SOXX & IDU  & IYF \\ \hline
\end{tabular}\label{tab:6}
\end{adjustwidth}
\end{table}

\begin{table}[h!]
\begin{adjustwidth}{-0.4in}{0in}
\caption{ \bf Top three US ETFs in each sub-period that transmit their shocks to the highest number of cryptocurrencies}
\begin{tabular}{cccccccc}
\hline
Influential Order & Pre-Covid-19 & Covid-19 & Bull Time 1 & Bull Time 2 & Bull Time 3 & U-R Conflict 1 & U-R Conflict 2 \\ \hline
1 & QYLD & SPY & IVE  & SLYG & IJS & QQQ & FDLO  \\
2 & SVXY & OEF & IYY & SVXY & FTXO  & SPY  & DVY \\
3 & IWB  & IUSG & KBWB & IYE  & KBWB & IJK  & IJH \\ \hline
\end{tabular}\label{tab:7}
\end{adjustwidth}
\end{table}

\section{Conclusions} \label{sec:5}

This work aims to detect and analyze investor herding behavior in different types of investment vehicles separately and also in their combination, including stocks, US ETFs and cryptocurrencies. We enrich the existing literature with three contributions: Firstly, the use of a more up-to-date and granular historical price time series covering several recent events, i.e. the US-China trade war in 2019, the Covid-19 pandemic followed by the global economic crisis in 2020, the Bull market period 2021-2022 and the Ukraine-Russia conflict which started in 2022, provides the latest insights into herding behavior in financial markets at a finer scale (30-min granularity). Secondly, the use of multiple investment vehicle types, enables us to examine herding behavior within a wider market context. Thirdly, the focus on herding behavior at a community level, enhances the clarity and granularity of this herding phenomenon.

Regarding the first research question, when examining the herding behavior in each type of investment vehicle separately, we find similar herding patterns between stocks and US ETFs for the majority of the time considered, with evidence of herding in both during the Pre-Covid-19 coinciding with the US-China trade war and during the Covid-19 pandemic. By contrast, the cryptocurrency market displays quite distinct behavior since its investors are seen to have herded during the Covid-19 pandemic, the first Bull Time when the market was recovering, the second Bull Time when most assets experienced a surge in prices and also the second Ukraine-Russia Conflict when another upward trend in financial markets took place.

Regarding the second research question, we execute a deeper herding detection and analysis by breaking down the whole combination of stocks, cryptocurrencies and US ETFs)
into smaller communities where each of these comprises the assets that are most similar to each other. Additionally, for each community, we also use information about sector distributions, with each asset belonging to one of the 13 sectors, including cryptocurrency, US ETF and 11 stock sectors. Our main finding is that herding behavior exists at all times but at a smaller scale (i.e. community level), in all types of investment vehicle examined. This mainly comes from specific events which particularly impact on that community. As a result, although herding might not be seen in a whole market composed of all assets, it can be found by looking at a particular community/subset of assets. Moreover, we notice 4 stock sectors in which for each sector, its assets and relevant ETFs tend to form communities among themselves, with the herding behavior found in these communities driven by unique events only influencing that sector. These include Energy, Healthcare, Information Technology and Financials. On the other hand, this characteristic does not appear in the remaining sectors, i.e. Industry, Consumer Discretionary and Consumer Staples.

This work considerably adds to the body of knowledge around understanding herding tendencies in the examined types of investment vehicle, principally to help investors construct an appropriate investment strategy. Apart from several factors causing herding behavior within a market such as investor sentiment, macroeconomics, major impactful events, news, and internal price movements,  US ETFs seem to be another herding driver in financial markets, especially in the direction of cryptocurrencies. This is reflected in the significant contagion effects that US ETFs exert on both stocks and cryptocurrencies. Additionally, contagion from stocks to US ETFs is consistently stronger than from cryptocurrencies to US ETFs. This tighter connection between US ETFs and stocks explains why they tend to form communities together. Furthermore,  we also raise investors' attention towards 4 stock sectors, i.e. Energy, Healthcare, Information Technology and Financials as these seem more sensitive to herding behavior than other sectors such as Industry, Consumer Discretionary and Consumer Staples. Therefore, investors who invest in these former four stock sectors should be more cautious with relevant economic and political turmoil since they are more likely to form a herding behavior during such occasions. Future work could expand on this study by using a larger dataset, covering more types of investment vehicle, such as commodities, currencies and bonds. Another area of future research could involve different time granularities. Furthermore, several findings of this study warrant further investigation: Firstly, the differences in herding magnitude between different communities. Secondly, the comparison between CSAD, CSAD for \textit{up} and CSAD for \textit{down} markets within each community, between different communities and between different sub-periods. Thirdly, the connections between ETFs tracking the S$\&$P-related indices and technology stocks.

\section*{Supporting Information}

\paragraph*{S1 Appendix.}
\label{SI_Appendix1}
{\bf The process of community detection.}

\paragraph*{S1 Text.}
\label{SI_Text2}
{\bf Visualization of herding detection results and sector distribution.} 

\paragraph*{S2 Text.}
\label{SI_Text3}
{\bf Time-varying community structures of cryptocurrencies, stocks and US ETFs.} 

\paragraph*{S1 Table.} 
\label{S1_Tab}
{\bf List of Assets Used in this Study.}

\paragraph*{S2 Table.}
\label{S2_Tab}
{\bf Classification of Stock Sectors in the Stock Markets.}

\paragraph*{S3 Table.}
\label{S3_Tab}
{\bf Details of Assets in each Community (Entire Period).}

\paragraph*{S4 Table.}
\label{S4_Tab}
{\bf Details of Assets in each Community (Sub-Period).}

\paragraph*{S1 File.}
\label{S1_File}
{\bf Contagion Effects between Cryptocurrencies, Stocks and US ETFs in each Sub-period.}


\end{document}